\documentclass{jfm}

\usepackage{graphicx}
\graphicspath{{Images/}}
\usepackage{epstopdf, epsfig}
\usepackage{subcaption}
\usepackage{comment}
\usepackage{amsmath, amsfonts, amssymb}
\usepackage{mathtools}
\usepackage{float}
\usepackage{enumitem}
\usepackage{mathptmx}
\usepackage{xcolor}

\usepackage{graphicx}
\usepackage{newtxtext}
\usepackage{newtxmath}
\usepackage{natbib}
\usepackage{hyperref}
\hypersetup{
    colorlinks = true,
    urlcolor   = blue,
    citecolor  = black,
}

\newcommand{\RomanNumeralCaps}[1]
\linenumbers

\title{Buoyancy effects on film boiling heat transfer from a sphere at low velocities.}

\author{Rishabh Singh,
  Anikesh Pal
  \corresp{\email{pala@iitk.ac.in}},
  \and Santanu De}
 
\affiliation{Department of Mechanical Engineering, Indian Institute of Technology Kanpur, INDIA}

\begin{document}
\maketitle

\begin{abstract}
A theoretical model is developed for the forced convection film boiling phenomenon over a heated sphere moving vertically downwards in the water. Unprecedented to the previous analytical studies, this model accounts for the buoyancy effects while solving the momentum, and energy equations in the vapor phase to obtain the velocity, and the temperature distribution in terms of the vapor boundary layer thickness. To calculate the vapor boundary layer thickness an energy balance is applied at the vapor-liquid interface. The flow of liquid around the sphere is considered to be governed
by the potential theory, and the energy equation in liquid is then solved for the known velocity distribution. We find that the vapor boundary layer thickness increases with an increase in the sphere, and the bulk water temperature, and a decrease in the free stream velocity. This further results in a decrease in the film boiling heat transfer coefficient. The present study concludes that at low free stream velocities ($< 0.5$
m/s) buoyancy becomes significant in delaying the separation, and when the velocity is further reduced the separation angle approaches $180^{\circ}$. 
\end{abstract}



\section{Introduction}
The knowledge of heat-transfer rates from spherical particles at high flux levels can significantly contribute towards designing energy systems associated with space industries and nuclear reactors. The primary mode of heat transfer in such energy systems is film boiling in which a vapor layer wraps the heated spherical surface preventing its contact with the liquid.
Film boiling can be characterized as natural convection film boiling and forced convection film boiling. In natural convection film boiling the motion of the liquid over the heated specimen is caused by the viscous drag forces of the rising vapor acting on the liquid whereas in forced convection film boiling the liquid is forced to flow over the heated specimen. 
The information about the film boiling phenomenon can be used to determine core cooling-ability after certain hypothetical nuclear accidents that result in extensive core melting. The concept of film boiling has also been utilized in the area of naval applications for drag reduction techniques by inserting a vapor layer in between the surface, and the surrounding liquid \citep{vakarelski2011drag}.\\

Theoretical and experimental investigation on film boiling of saturated liquid such as carbon tetrachloride, benzene, ethyl alcohol, n-hexane over a cylinder were performed by \cite{bromley1953heat}. 
They reported that for high velocity flows the separation angle was close to $90^\circ$, whereas for sufficiently low velocities the separation angle approaches 180$^\circ$. \cite{motte1957film} used the same experimental setup as \cite{bromley1953heat} with some modifications to study subcooled (when the temperature of the liquid is below its boiling point) forced convection film boiling over a cylinder with turbulence. It was found that with an increase in subcooling and velocity, the heat transfer rate also increases. \cite{bradfield1967effect} also studied film boiling over a sphere using experimental techniques, and concluded that minimum superheat required to sustain the film boiling increases linearly with an increase in subcooling. Transient subcooled forced convection film boiling over a sphere was experimentally investigated by \cite{walford1969transient}. Different regimes of film boiling over the sphere have been identified, and the subsequent heat flux behavior in those regimes was reported. \\

\cite{kobayasi1965film} theoretically investigated film boiling heat transfer from a sphere moving downward in a liquid and proposed a general solution for predicting the boiling heat transfer coefficient as a function of certain parameters such as Reynolds number, liquid-vapor viscosity ratio, Prandtl number, size of the sphere, the kinematic viscosity of the liquid. However, the findings of \cite{kobayasi1965film} were not accurate owing to the incorrect pressure used for the theoretical derivation \citep{hesson1966comment}.\\

To derive the theoretical heat transfer rate \cite{bromley1953heat, kobayasi1965film} used imposed pressure gradient from the free stream. Additionally, Bernoulli’s theorem was applied to get an additional equation in terms of the frictional loss in vapor. The problem was further simplified by considering saturated liquid flow around the body. When the liquid is at saturation temperature there will be no heat flux going into the bulk liquid, and all the heat leaving the sphere is used in vaporizing and superheating the vapor. As the heat transfer phenomenon is straightforward in the case of saturated liquid, the energy conservation equation is not solved, and the calculations in the liquid become simple. However, in practical situations, the liquids are not saturated. Therefore, an energy equation should be solved both in the liquid and the vapor phase to obtain an accurate temperature distribution to properly characterize the heat transfer process around the body. In the present investigation, we solve the energy equation in both the liquid and the vapor phases to obtain the temperature distribution in both phases.\\

\cite{Witte67,witte1968a,witte1968b, Witte1984} carried out experimental and theoretical investigations of forced convection film boiling from a sphere moving in a liquid. The experiment of \cite{witte1968a} used a transient technique in which a heated sphere attached to a swinging-arm apparatus was passed through a pool of liquid sodium. The heat transfer rates from the sphere to liquid sodium were measured, and were found to be in good agreement with the theoretical expressions for heat transfer from a sphere during forced convection with the assumption of potential flow in liquid sodium. \cite{witte1968b} assumed a linear profile for velocity in the vapor film and reported the forced convection film boiling from a sphere in a saturated liquid. The effect of non-linear velocity profile within the vapor film on subcooled flow film boiling from a sphere is analyzed by \cite{Witte67, Witte1984}. While calculating the vapor boundary layer thickness \cite{Witte67} neglected the effect of radiation, and argued that for highly subcooled liquid the energy required for the vaporization of liquid can be ignored in comparison to the heat energy going into the bulk liquid. In contrast, \cite{Witte1984} included the heat energy required to vaporize the liquid and concluded that the results based on the non-linear velocity profile produce results comparable to the experiments. The liquid velocity at the vapor-liquid interface was calculated from the potential flow theory in all the investigations. Additionally these theoretical investigations did not consider buoyancy in their analysis. We will demonstrate in section \ref{results} that buoyancy plays a crucial role in obtaining results that are similar to the experiments.\\

\cite{Dhir78} performed theoretical and experimental investigation to determine the effect of flow velocity, subcooling, initial sphere temperature on film boiling heat transfer from a sphere. Their theoretical analysis although included the effect of buoyancy was restricted only to natural convection film boiling over a sphere where the surrounding liquid was stagnant. The vapor film was assumed to be stable, and very thin in comparison to the radius of the sphere so that the non-linear behavior of the film can be neglected. With an increase in both sphere, and bulk water temperature  \cite{Dhir78} observed a decrease in the film boiling heat transfer coefficient owing to an increase in the vapor film thickness. They also reported that the minimum temperature to sustain a stable film depends only on subcooling, and increases linearly with subcooling.\\ 

An experimental study of transient film boiling on different geometries (spheres, cylinders, flat plates) with different coolant velocities was also conducted by \cite{Jouhara09}. Their study on the nature of the vapor/liquid interface and the collapse modes has revealed a new model for film collapse, in which an explosive liquid-solid contact is followed by film re-formation and the motion of a quench front over the hot surface. The heat transfer coefficients, and heat fluxes during film boiling were found essentially to depend on the temperature of the body, and water subcooling. A theoretical model was also developed that predicted the heat transfer coefficients to within $10\%$ of experimental values for water subcooling above $10K$. However, their theoretical model was restricted to plane surfaces only.\\

In this investigation, we develop a theoretical model to determine the heat transfer characteristics and boundary layer separation behavior owing to film boiling from a heated spherical particle moving slowly in water under the influence of buoyancy unprecedented to the earlier theoretical studies. A comparison of our theoretical model with the experimental study of \cite{Jouhara09}, and the theoretical model of \cite{Witte1984} is also performed to access the efficacy of our model.\\

The methodology for the development of the theoretical model is presented in section \ref{methods}. Results from our model are discussed in section \ref{results} and the conclusions drawn from this study are given in section \ref{summary}.\\

\section{Methodology}
\label{methods}
The schematic of film boiling over a sphere is shown in figure \ref{fallingsphere}. When the liquid comes in contact with the heated sphere, a vapor layer is formed around the sphere as the temperature of the sphere is higher than the saturation temperature of the liquid. Heat conduction occurs through the vapor layer. A portion of this heat is utilized in vaporizing the liquid that adds to the vapor layer, increasing the vapor layer thickness. Another portion of the heat is diffused into the bulk liquid. Figure \ref{fallingsphere} manifests the vapor layer and the liquid layer around the sphere. The vapor boundary layer moves past the heated sphere and is influenced by both the sphere and the liquid layer. The liquid layer only feels the influence of the vapor layer and is not in direct contact with the sphere. We aim to theoretically determine the heat transfer rates during the film boiling from the sphere including the effects of buoyancy. Our analysis is based on the following assumptions:\\
\begin{enumerate}
    \item \,\, liquid-vapor interface is smooth and is in dynamic equilibrium, 
    
    \item \,\, the temperature of the sphere is uniform,
    
    \item \,\, physical properties of vapor and liquid are evaluated at mean film temperature,
    
    \item \,\, heat transfer across the vapor layertake place by conduction only,
    
    \item \,\, inertial effects in the momentum and energy equations are neglected,  
    
    \item \,\, the flow of liquid around the sphere is governed by potential flow theory,
    
    \item \,\, vapor film is axially symmetric.\\

\end{enumerate}

All of the above-mentioned assumptions are justified from the available theoretical and experimental studies. \cite{Bradfield66} observed that the ripples formed during film boiling in the surrounding liquid at rest tends to dampen as the liquid starts moving. As the velocity of the liquid around the body is increased the liquid-vapor interface becomes unstable. The velocity range we use in the current investigation is smaller than that can cause an unstable interface and therefore, it is reasonable to assume a smooth liquid-vapor interface. The uniformity of the temperature of the sphere is justified for low Biot numbers.
\cite{bradfield1967effect} also found that the maximum discrepancy in heat flux calculations was less than $2\%$ if uniform temperature distribution is assumed within the specimen as compared to the case when calculations are performed considering variability in temperature distribution within the sphere. The physical properties of the vapor and the liquid phase were computed by \cite{bromley1953heat} from well-defined expressions developed to calculate the average value of the physical property. However, it was concluded that for simplicity all physical properties can be evaluated at the mean film temperature except for the latent heat of vaporization. This justifies our third assumption. \cite{Burns89} concluded that the film thickness obtained experimentally, and calculated assuming heat transfer across the film by conduction manifest no significant difference. Therefore, the heat transfer across the vapor film can be assumed to take place solely by conduction. Similar assumptions were made by \cite{bromley1953heat,motte1957film}. Our fifth assumption is justified owing to the fact that the thickness of the vapor layer is very small in comparison to the diameter of the sphere \citep{bromley1953heat, Witte67,kobayasi1965film, Witte1984, Jouhara09}.
Assumption 6 is justified from the study of \cite{Kutateladze59} where it has been shown that the assumption of potential flow or viscous flow in the liquid does not make a significant difference. 

\subsection{\textbf{Liquid region}} \label{liqregion}
According to the sixth assumption, the velocity distribution in bulk liquid is:
\begin{equation}
    u_r=-3U\frac{r-R}{R}\cos\theta \hspace{1cm} , \hspace{1cm}  u_\theta=\frac{3}{2} U\sin\theta,
\end{equation}
where, $\theta$ is the azimuthal angle measured from the stagnation point, $r$ is the radial direction, $U$ is the incoming free stream velocity of liquid, $u_r$ is the velocity in the radial direction, $u_\theta$ is the velocity in the azimuthal direction, $R$ is the radius of the sphere. \\

\begin{figure}
    \begin{center}
      \includegraphics[scale=0.35]{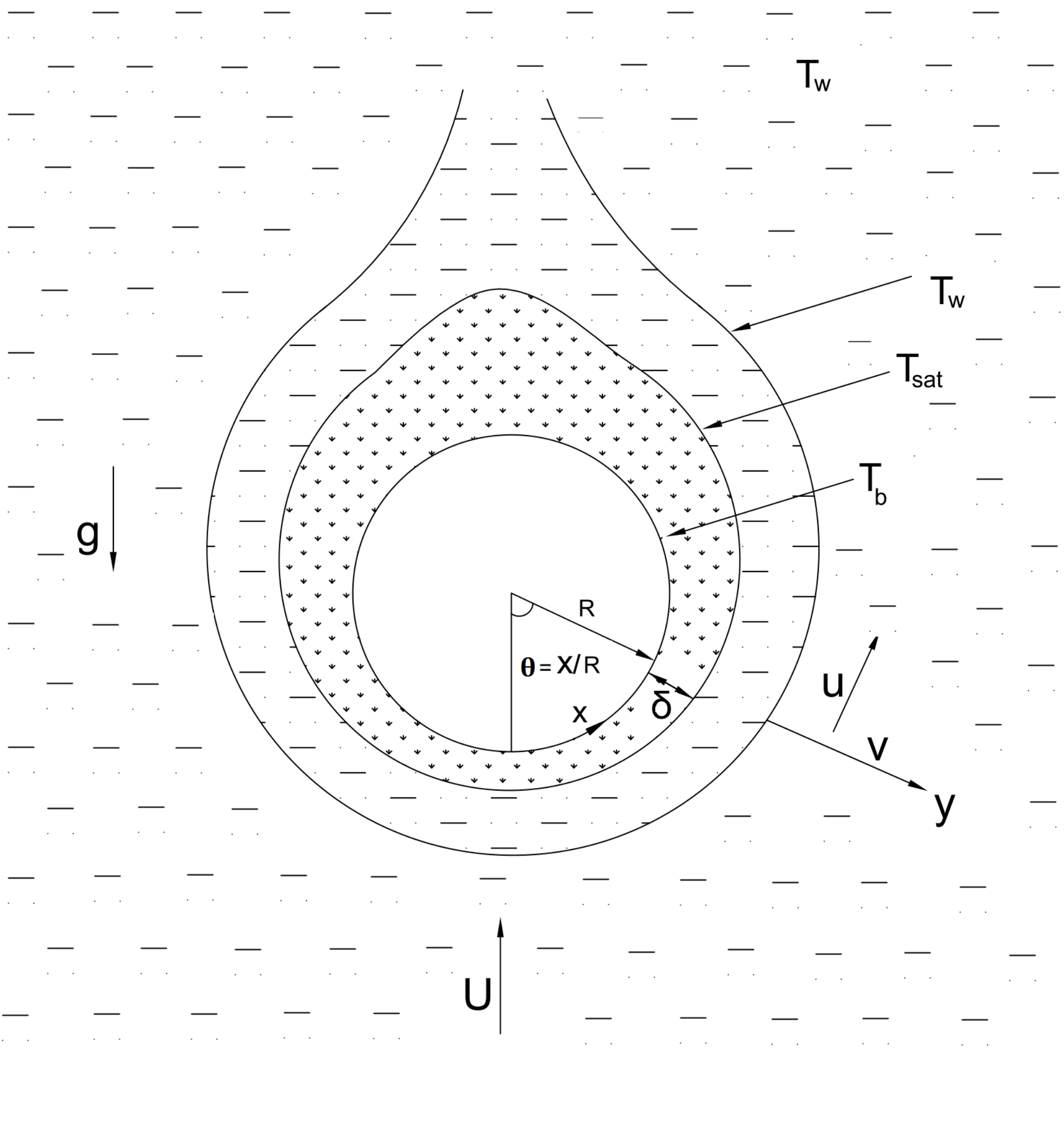}
    \end{center}
    \caption{Schematic of film boiling over a sphere.  \protect} 
    \label{fallingsphere}
\end{figure}

From figure \ref{fallingsphere}, we can write:\\
\begin{enumerate}
\item\,\, $y=r-R$,  
\item\,\, $\theta=\frac{x}{R}$,\\
\end{enumerate} 
where $x$ is the curvilinear coordinate along the surface of the sphere, and $y$ is the curvilinear coordinate normal to the $x$ direction. \\

We can transform $u_r$ and $u_\theta$ in curvilinear coordinate system
as follows:\\
\begin{equation}
u_r=-3U\frac{r-R}{R}\cos\theta = -3U\frac{y}{R}\cos\frac{x}{R}, \label{eqvinliq}
\end{equation}

\begin{equation}
u_\theta=\frac{3}{2} U \sin \theta = \frac{3}{2} U \sin \frac{x}{R},
\label{equinliq}
\end{equation}

Next, we consider the energy equation for the liquid in the spherical coordinates system:\\
\begin{equation}
u_r\frac{\partial{T}}{\partial{r}} + \frac{u_\theta}{r}\frac{\partial{T}}{\partial{\theta}} +\frac{u_\phi}{r\sin\theta} \frac{\partial{T}}{\partial{\phi}} = \alpha_l\left(\frac{\partial^2{T}}{\partial{r^2}} + \frac{2}{r} \frac{\partial{T}}{\partial{r}} \right).
\label{eqsph}
\end{equation}

where $\alpha_l$ is the thermal diffusivity of liquid and $T$ is temperature. Since the flow is assumed to be axially symmetric and there is no component of velocity in the $\phi$ direction, we can write equation \ref{eqsph} as follows:\\

\begin{equation}
u_r\frac{\partial{T}}{\partial{r}} + \frac{u_\theta}{r}\frac{\partial{T}}{\partial{\theta}}  = \alpha_l\left(\frac{\partial^2{T}}{\partial{r^2}} + \frac{2}{r} \frac{\partial{T}}{\partial{r}} \right). 
\label{momeqliqsph}
\end{equation}

\cite{Sideman66} demonstrated that if heat transfer is assumed to take place in a thin layer near the interface, the term scaling with $ \frac{1}{r} \frac{\partial{T}}{\partial{r}} $ can be neglected in comparison to the term $ \frac{\partial^2{T}}{\partial{r^2}} $. Therefore, under this assumption we can modify the equation \ref{momeqliqsph} as follows:

\begin{equation}
u_r\frac{\partial{T}}{\partial{r}} + \frac{u_\theta}{r}\frac{\partial{T}}{\partial{\theta}}  = \alpha_l\frac{\partial^2{T}}{\partial{r^2}}.
\label{eqliq}
\end{equation}

We use the information from figure \ref{fallingsphere} for the following transformations:
\begin{equation}
u_\theta =u_l,\,\, u_r =v_l \,\,;\,\, x= r\theta \implies dx = rd\theta\,\, ;\,\, y = r-R \implies  dy= dr,
\label{T}
\end{equation}

where, $u_l$ and $v_l$ are the velocities of the liquid in $x$ and $y$ directions respectively. 
Substituting \ref{T} in equation \ref{eqliq} we get,

\begin{equation}
u_l\frac{\partial{T}}{\partial{x}} + v_l\frac{\partial{T}}{\partial{y}} = \alpha_l\frac{\partial^2{T}}{\partial{y^2}}
\label{goveqliq}
\end{equation}

The boundary conditions corresponding to equation \ref{goveqliq} considering $R+\delta \sim R$
are as follows:\\
\begin{enumerate}
\item\,\, $ y \to \infty \,\, , \,\, T=T_w \,\, , \,\, \theta\geq 0 $,
\item\,\, $y = 0  \,\,  ,  \,\,  T=T_{sat}, \,\, \theta \geq 0 $, 
\item\,\, $ 0<   \,\, y \leq \infty \,\, , \,\, T=T_w, \,\, \theta = 0 $.\\
\end{enumerate}

Here, $T_{sat}$ is the saturation temperature of the liquid, $T_w$ is the temperature of bulk water, $\delta$ is the vapor layer thickness. We transform equation \ref{goveqliq} using the following variables such that the solution of the transformed equations is known. 

\begin{equation}
\Delta T= T- T_{sat},\,\ \psi=y\sin^2 \theta ,\,\,  \eta=\int \limits_{0}^{\theta}\sin^3\theta d\theta = -\frac{3}{4}\cos\theta+ \frac{1}{12}\cos3\theta + \frac{2}{3}. 
\end{equation}
Now, consider the following derivatives

\begin{equation}
\frac{\partial{T}}{\partial{y}} = \frac{\partial{\Delta T}}{\partial{y}}=
\frac{\partial{\Delta T}}{\partial{\psi}}\frac{\partial{\psi}}{\partial{y}}\,\, + \,\, \frac{\partial{\Delta T}}{\partial{\eta}}\frac{\partial{\eta}}{\partial{y}}\,\, = \,\, \frac{\partial{\Delta T}}{\partial{\psi}}\sin^2 \theta.
\label{eqi}
\end{equation}

\begin{equation}
\frac{\partial^2{T}}{\partial{y}^2} = \frac{\partial^2{\Delta T}}{\partial{y}^2} = \frac{\partial}{\partial{y}}\left(  \frac{\partial{\Delta T}}{\partial{y}} \right) = \frac{\partial{}}{\partial{\psi}} \left(\frac{\partial{\Delta T}}{\partial{\psi}}\sin^2 \theta \right)\frac{\partial{\psi}}{\partial{y}} + \frac{\partial{}}{\partial{\eta}} \left(\frac{\partial{\Delta T}}{\partial{\psi}}\sin^2 \theta \right)\frac{\partial{\eta}}{\partial{y}} = \sin^4 \theta \frac{\partial^2{\Delta T}}{\partial{\psi^2}}
\label{eqii}.
\end{equation}

\begin{equation}
\frac{\partial{T}}{\partial{x}} = \frac{\partial{\Delta T}}{\partial{x}} = \frac{\partial{\Delta T}}{\partial{\psi}}\frac{\partial{\psi}}{\partial{x}}\,\, + \,\, \frac{\partial{\Delta T}}{\partial{\eta}}\frac{\partial{\eta}}{\partial{x}}\,\, = 2\frac{y}{R}\sin\theta\cos\theta\frac{\partial{\Delta T}}{\partial{\psi}}+ \left(\frac{3}{4R}\sin\theta - \frac{1}{4R}\sin3\theta \right)\frac{\partial{\Delta T}}{\partial{\eta}}. 
\label{eqiii}
\end{equation} 

Substituting equations \ref{eqvinliq}, \ref{equinliq}, \ref{eqi}, \ref{eqii}, \ref{eqiii} in equation \ref{goveqliq}, we get:\\ 
 
\begin{equation}
\frac{\partial{\Delta T}}{\partial{\eta}}\, = \, \frac{2R\alpha_l}{3U}\,\,\frac{\partial^2{\Delta T}}{\partial \psi^2}.
\end{equation}

Using ${M}= \frac{2R\alpha_l}{3U} $, we can write:

\begin{equation}
\frac{\partial{\Delta T}}{\partial{\eta}} =  M \frac{\partial^2{\Delta T}}{\partial \psi^2}.
\label{eqgovtrans}
\end{equation}

The boundary conditions corresponding to equation \ref{eqgovtrans} are:\\

\begin{enumerate}
\item\,\, $ \psi \to \infty \,\, , \,\, \eta \geq 0 \,\, , \,\, \Delta T= T_w-T_{sat} $,
\item\,\, $ \psi=0 \,\, , \,\, \eta \geq 0 \,\, , \,\, \Delta T =0 $,
\item\,\, $ 0<   \,\, \psi \leq \infty \,\, , \,\, \eta=0 \,\, , \,\,     \Delta T= T_w-T_{sat} $.\\
\end{enumerate}

Solution of equation \ref{eqgovtrans} subjected to the above boundary conditions can be found by defining $\beta =\frac{T-T_w}{T_{sat}-T_{w}}$ and using it in equation \ref{eqgovtrans} to get: 

\begin{equation}
\frac{\partial{\beta}}{\partial{\eta}}\, = \, M \,\,\frac{\partial^2{\beta}}{\partial \psi^2}. 
\label{modgoveqliq}
\end{equation}

The boundary conditions corresponding to equation \ref{modgoveqliq} are:\\
\begin{enumerate} 
\item \,\, $ \psi \to \infty \,\, , \,\, \eta \geq 0 \,\, , \,\, \beta=0 $,
\item\,\, $ \psi=0 \,\, , \,\, \eta \geq 0 \,\, , \,\, \beta =1 $,
\item\,\,  $ 0<   \,\, \psi \leq \infty \,\, , \,\, \eta=0 \,\, , \,\, \beta=0 $.\\
\end{enumerate}

The partial differential equation \ref{modgoveqliq} can be converted to ordinary differential equation using method of combination of variable.
Defining $\beta = \psi^a\, \eta^b $, where a and b are constants and substituting in equation \ref{modgoveqliq} we get: 

\begin{equation}
\frac{\psi^2}{M\eta} = \frac{a(a-1)}{b} =\mathrm{Constant}
\end{equation}

The new variable can be of the form $\left(\frac{c \psi^2}{\mathrm{M}\eta}  \right)^d $.
Let us define the new variable as $\gamma = \frac{\psi}{\sqrt{4M\eta}}$ (obtained by choosing d$=1/2$ and c$=1/4$) and hence we can write  $\beta(\psi, \eta)= \beta(\gamma)$.

\begin{equation}
\frac{\partial{\beta}}{\partial{\eta}} = \frac{d\beta}{d\gamma}\frac{\partial{\gamma}}{\partial{\eta}} = - \frac{\psi}{2\eta\sqrt{4\eta M}}  \frac{d\beta}{d\gamma}
\label{above1}
\end{equation}

\begin{equation}
\frac{\partial^2{\beta}}{\partial{\psi^2}}= \frac{1}{4 M \eta}\frac{d^2\beta}{d\gamma^2}
\label{above2}
\end{equation}

Substituting equation \ref{above1} and \ref{above2} in equation \ref{modgoveqliq}, we get an ordinary differential equation as follows:

\begin{equation}
 \frac{d^2 \beta}{d\gamma^2} + 2\gamma\frac{d\beta}{d\gamma} =0.
 \label{goveqbeta}
\end{equation}

The boundary conditions corresponding to equation \ref{goveqbeta} will become:\\
\begin{enumerate}
\item\,\, $\gamma=0 \,\, , \,\, \beta=1 $,
\item\,\, $\gamma=\infty \,\, , \,\, \beta=0  $.\\
\end{enumerate}

The solution of equation \ref{goveqbeta} is of the form:\\

\begin{equation}
\beta= B + A \int\limits_{0}^{\gamma}e^{-\gamma^2} d\gamma,
\end{equation}

and applying the boundary conditions will result in:\\
\begin{equation}
\frac{T-T_w}{T_{sat}-T_w} = erf_c\left(\frac{\psi}{2\sqrt{\mathrm{M}\eta}}. \right)
\label{expTliq}
\end{equation}

Equation \ref{expTliq} represents the temperature distribution in the liquid, and we can use it to calculate the heat flux, $q^{\prime\prime}_b$, into the bulk liquid as follows:

\begin{equation}
q^{\prime\prime}_b=-k_l\left.\left(\frac{\partial{T}}{\partial{y}}\right)\right |_{y=0},
\end{equation}
where $k_l$ is the thermal conductivity of the liquid.

Using equations \ref{eqi} and \ref{expTliq} we get, 

\begin{equation}
\left.\frac{\partial{T}}{\partial{y}}\right|_{y=0} = -\frac{(T_{sat}-T_w)\sin^2\theta}{\sqrt{\pi M\eta}}, 
\end{equation}

\begin{equation}
q''_b= \frac{k_l\Delta T_w \sin ^2 \theta}{\sqrt{\pi M\eta}}.
\end{equation}

\subsection{\textbf{Vapor region}} \label{vapor region}

We write the momentum equation in the $x$ direction in the vapor region following figure \ref{fallingsphere} as follows:\\
\begin{equation}
\rho_v \left(u\frac{\partial{u}}{\partial{x}} + v\frac{\partial{u}}{\partial{y}}\right)= -\frac{\partial{p}}{\partial{x}} + \Delta \rho g \sin\theta  + \mu_v \frac{\partial^2{u}}{\partial{y^2}},
\label{momeqvap}
\end{equation}\\

where $\Delta \rho= \rho_l-\rho_v$, $\rho_l$ and $\rho_v$ are the densities of liquid and vapor respectively and $g$ is the acceleration due to gravity, $u$ and $v$ are the velocities of the vapor in $x$ and $y$ directions respectively. Application of the fifth assumption results in the following equation:

\begin{equation}
\frac{\partial^2{u}}{\partial{y^2}} = \frac{1}{\mu_v}\left(\frac{\partial{p}}{\partial{x}} - \Delta \rho g \sin\theta\right).
\label{momeqvap2}
\end{equation}

The boundary conditions corresponding to equation \ref{momeqvap2} are,
\begin{enumerate}
\item\,\, $ y=0 \,\, , \,\, u=0 $,
\item\,\, $ y=\delta \,\, , \,\, u=\frac{3}{2}U \sin\theta$.\\
\end{enumerate}

Since the vapor layer thickness $(\delta)$ is thin, the streamwise variation of pressure in the liquid layer as given by the Bernoulli equation is impressed on the vapor layer \citep{Witte67}. Therefore, using Bernoulli's equation in the liquid layer we can write: 

\begin{equation}
    p + \frac{1}{2}\rho_l u_l^2 = Constant.
    \label{Beq}
\end{equation}
Here $u_l$ is the velocity in the liquid. Differentiating equation \ref{Beq} with respect to $x$ we obtain:
\begin{equation}
    \frac{\partial{p}}{\partial{x}}=-\rho_l u_l \frac{d u_l}{dx}.
\label{bernoulli}    
\end{equation}

From equations \ref{eqvinliq} and \ref{T} we can write $u_l=\frac{3}{2} U \sin \theta = \frac{3}{2} U \sin \frac{x}{R}$ and modify equation \ref{bernoulli} as follows:
\begin{equation} 
\frac{dp}{dx}=- \rho_l u_l \frac{d u_l}{dx}=-\frac{9}{8}\left(\frac{\rho_l U^2}{R} \right)\sin 2\theta .
\label{momeqvap3}
\end{equation}

Substituting equation \ref{momeqvap3} in equation \ref{momeqvap2} and solving for the corresponding boundary conditions we get,

\begin{equation}
 u = \frac{3}{2} U\sin\theta \frac{y}{\delta} + \left( \frac{9}{8}\frac{\rho_l U^2}{\mu_v R} \sin\theta \cos\theta +  \frac{\Delta\rho g \sin \theta}{2\mu_v} \right)\left( y\delta - y^2 \right)
 \label{velvap}
 \end{equation}

We can see that the velocity in equation \ref{velvap} is comprised of a linear term, $\frac{3}{2} U\sin\theta \frac{y}{\delta}$, and two non-linear terms, $\frac{9}{8}\frac{\rho_l U^2}{\mu_v R} \sin\theta \cos\theta$, and $\frac{\Delta \rho g \sin \theta}{2\mu_v}$. The first non-linear is due to the imposed pressure gradient by the potential flow of liquid, whereas the second non-linear term represents the effect of buoyancy. \cite{Witte1984} in their theoretical model did not consider buoyancy effects. Therefore, if we neglect the buoyancy, then non-linearity in the velocity profile is sustained only by the imposed pressure. We can further see that the first non-linear term is proportional to the square of the velocity, and at low velocities, the non linear term is dominated by the buoyancy effects.\\

\subsection{\textbf{Temperature distribution in Vapor layer}}
In equation \ref{velvap} the vapor layer thickness, $\delta$, is an unknown. Determination of $\delta$ is important for understanding the heat transfer phenomenon. To compute $\delta$ we start with the energy equation for the vapor layer in the $x$ direction as follows:
 
\begin{equation}
u\frac{\partial{T}}{\partial{x}} + v\frac{\partial{T}}{\partial{y}} = \alpha \frac{\partial^2{T}}{\partial{y^2}}.
\label{eqTvap1}
\end{equation}

The corresponding boundary conditions for \ref{eqTvap1} are:
\begin{enumerate}
\item\,\, $ y=0 \,\, , \,\, u=0 \,\, ,  \,\, T=T_b $,
\item\,\, $ y=\delta \,\, , \,\,  u=\frac{3}{2}U \sin\theta \,\, ,  \,\, T=T_{sat}$,
\end{enumerate}

where $T_b$ is the temperature of the sphere. Using assumptions $4$ and $5$, equation \ref{eqTvap1} can be written, and solved as follows:

\begin{equation}
\frac{\partial^{2} T}{\partial{y^2}} = 0 \implies T=C_1y +C_2.
\label{eqTvap2}
\end{equation}

Substituting the corresponding boundary condition in equation \ref{eqTvap2} we get,
\begin{equation}
T= T_b + \left(T_{sat} - T_b \right) \frac{y}{\delta}.
\label{Tvapor}
\end{equation}

This equation represents the temperature distribution in the vapor layer.\\

\subsection{\textbf{Vapor boundary layer thickness}}
\begin{figure}
    \begin{center}
      \includegraphics[scale=0.25]{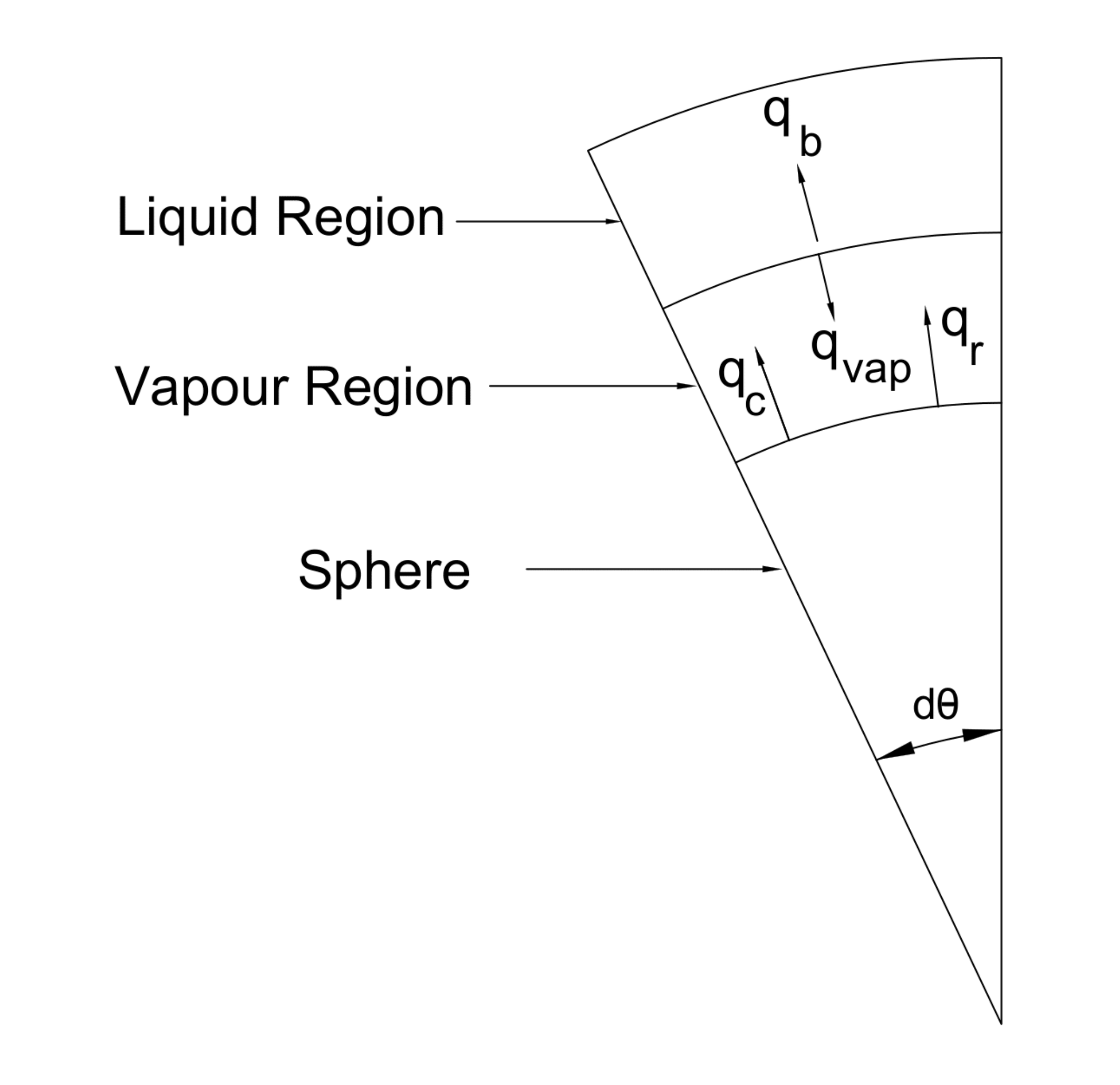}
    \end{center}
    \caption{Energy balance over elemental area of sphere.  \protect} 
    \label{energybal1}
\end{figure}
Next, we consider the heating provided by the sphere that results in the vaporization of liquid at the vapor-liquid interface, and superheating of the newly formed vapor above $T_{sat}$. Also, since the bulk water is below the saturation temperature, a part of total heat energy available at the vapor-liquid interface due to conduction across the vapor film, and radiation from the sphere goes into the bulk liquid. From the energy balance on a differential element as shown in figure \ref{energybal1} we can write, 
\begin{equation}
dq_c + dq_r =dq_{vap} + dq_b,
\label{energybal2}
\end{equation}

where, 
\begin{enumerate}
\item\,\, $dq_c$ is heat transfer due to conduction across vapor film,  $q^{\prime \prime}_c= \frac{k_v(T_b-T_{sat})}{\delta} $ ( we get by substituting equation \ref{Tvapor} in the Fourier's law of heat conduction,
\item\,\, $dq_r$ is heat transfer due to radiation , $q^{\prime \prime}_r= \sigma\epsilon(T^4_b - T^4_{sat})$
\item\,\, $dq_{vap}$ is heat utilised in vaporizing the liquid.
\item\,\, $dq_b$ is sensible heat energy going in water,  $q^{\prime \prime}_b= \frac{k_l\Delta T_w \sin ^2 \theta}{\sqrt{\pi \mathrm{M}\eta}}$\\
\end{enumerate}

The energy flux utilised in vaporization of liquid can be written as,
\begin{equation}
dq_{vap}=h'_{fg}dw = h'_{fg}d(\rho_v A_c\overline{u}), 
\label{qvap}
\end{equation}

where dw is the increase of mass flow rate in vapor layer due to vaporization, $h_{fg}$ is the latent of vaporization, $h'_{fg}= h_{fg}\left(1 + \frac{0.4 C_{p_l} \left( T_b - T_{sat}\right)}{h_{fg}}   \right) $ is the modified latent heat of vaporization  \citep{bromley1953heat,Witte67, Witte1984} that accounts for the temperature variation in the vapor field and super heating of vapor above $T_{sat}$ , $A_c= 2\pi R\delta\sin\theta$ is the flow cross section of the film, and  $\overline{u}$ is the average vapor velocity at any $\theta$.\\ 

The average velocity in the vapor film is calculated as follows:
\begin{equation}
\overline{u}=\frac{1}{\delta}\int\limits_{0}^{\delta}u\,dy = \frac{1}{\delta} \int\limits_{0}^{\delta} \left( \frac{3}{2} U \sin\theta \,\frac{y}{\delta} + \left(\,\, \frac{9}{8}\frac{\rho_l U^2}{\mu_v R} \sin\theta \cos\theta \,\, + \,\, \frac{\Delta \rho g \sin \theta}{2\mu_v} \right)\left( y\delta - y^2 \right)\right) dy
\end{equation}

$\implies$
\begin{equation}
\overline{u} = \frac{3}{4}U\sin\theta\, +\, \frac{3 \rho_l\ U^2}{16 \mu_v R}\sin\theta\cos\theta \delta^2\, +\, \frac{\Delta \rho g\sin\theta}{12\mu_v}\delta^2 
\label{uavg}
\end{equation}

Using equation \ref{uavg} in equation \ref{qvap},

\begin{equation}
dq_{vap}=h'_{fg}\,\,d\left(\rho_v 2\pi R\delta\sin\theta \left( \frac{3}{4}U\sin\theta\, +\, \frac{3 \rho_l\ U^2}{16 \mu_v R}\sin\theta\cos\theta \delta^2\, +\, \frac{\Delta \rho_l g\sin\theta}{12\mu_v}\delta^2 \right) \right)
\end{equation}
$\implies$
\begin{equation}
dq_{vap}=h'_{fg}\,\, \frac{d}{d\theta} \left(\rho_v 2\pi R\delta\sin\theta \left( \frac{3}{4}U \sin\theta\, +\, \frac{3 \rho_l\ U^2}{16 \mu_v R}\sin\theta\cos\theta \delta^2\, +\, \frac{\Delta \rho_l g\sin\theta}{12\mu_v}\delta^2 \right) \right)d\theta 
\label{eqdqvap}
\end{equation}

From equation \ref{energybal2} we can write, 
\begin{equation}
\frac{k_v \left(T_b - T_{sat}\right)}{\delta}dA \,+ \, q''_r dA\, = \, dq_{vap} +q''_b dA, 
\end{equation}
$\implies$
$$ dq_{vap}= \frac{k_v  \left(T_b - T_{sat}\right)}{\delta}dA \,+ \, q''_r dA\,  \,  - q''_b dA $$,
$\implies$
$$ \frac{ dq_{vap}}{dA}= \frac{k_v  \left(T_b - T_{sat}\right)}{\delta} \,+ \, q''_r \,  \,  - q''_b  \hspace{2cm} $$,

where, $dA = 2\pi R^2 sin\theta d\theta$ is the differential area element on the sphere, and $k_v$ is the thermal conductivity of the vapor. Substituting, $dA$ , $q^{\prime \prime}_r$ and  $q^{\prime \prime}_b  $ in above equation, we get

\begin{equation}
\frac{ dq_{vap}}{2\pi R^2 sin\theta d\theta}= \frac{k_v  \left(T_b - T_{sat}\right)}{\delta} \,+ \, \sigma\epsilon(T^4_b - T^4_{sat}) \,  \,  - \frac{k_l\Delta T_w \sin ^2 \theta }{\sqrt{\pi \mathrm{M}\eta}}, 
\label{diffEbal}
\end{equation}

Substituting equation \ref{eqdqvap} in \ref{diffEbal} , and separating $\frac{d\delta}{d\theta}$ we obtain:

\begin{equation}
\frac{d\delta}{d\theta}= \frac{ \frac{k_v  \left(T_b - T_{sat}\right)}{\delta}\,+\, \sigma\epsilon(T^4_b - T^4_{sat})\,-\frac{k_l\Delta T_w \sin ^2 \theta }{\sqrt{\pi M\eta}}\, -\frac{h'_{fg}\rho_v}{R}\left(\frac{3U \cos\theta \delta}{2} + \frac{3 \rho_l U^2}{16\mu_v R}(3\cos^2\theta -1)\delta^3 + \frac{\Delta \rho g \cos\theta}{6\mu_v}\delta^3  \right)}{ \frac{h'_{fg}\rho_v}{R} \left( \frac{3 U \sin\theta}{4} + \frac{9\rho_l U^2}{16\mu_v R}\sin\theta \cos\theta\delta^2 + \frac{\Delta \rho g \sin\theta}{4 \mu_v}\delta^2  \right)}
\label{diffdelta}
\end{equation}

The non-dimensional form of equation \ref{diffdelta} is shown below (the steps to non-dimensionalize equation \ref{diffdelta} is given in the appendix).\\ 

\begin{equation}
\begin{aligned}
\frac{d(\frac{\delta}{D})}{d\theta} = \frac{1}{1+ \frac{3\rho_l}{2\rho_v}Re_v (\frac{\delta}{D})^2 \cos\theta + \frac{1}{3}(\frac{\delta}{D})^2 \frac{G_r}{Re_v}} \left(\frac{2 J_v}{3 Pe_v\sin\theta(\frac{\delta}{D})}+\frac{2 q_r}{3 \rho_v U h'_{fg} \sin\theta} -\right. \\ \\ \left. 2\left(\frac{\delta}{D}\right)\cot\theta-\frac{1}{2} \frac{\rho_l}{\rho_v}Re_v \left(\frac{\delta}{D}\right)^3(\frac{3\cos^2\theta -1}{\sin\theta})-
\frac{2}{9}\frac{Gr}{Re_v}\left(\frac{\delta}{D}\right)^3 \cot\theta - \right. \\ \frac{2 \frac{\rho_l}{\rho_v} J_l \sin\theta}{3\left( \frac{\pi Pe_l}{3}\left(\frac{2}{3}-\cos\theta + \frac{\cos^3\theta}{3}  \right) \right)^\frac{1}{2} }  \Bigg).
\label{eqNDdeltaWB1}
\end{aligned}
\end{equation}

Here, $Re_v=\frac{\rho_v U D}{\mu_v}$ is the vapor Reynolds number, $Gr=g \left(\frac{\rho_l}{\rho_v}-1\right) \frac{D^3}{\nu_v ^2}$ is the  Grashof number (representing the ratio of buoyancy force to the viscous force acting on a fluid), $J_v=\frac{C_{p_v} \left(T_b-T_{sat}\right)}{h'_{fg}}$ and $J_l=\frac{C_{p_l}\left(T_{sat}-T_w\right)}{h'_{fg}}$ are the vapor and liquid Jakob numbers respectively (representing the sensible heat absorbed or released during the liquid vapor phase change in comparison to the latent heat), $Pe_v=\frac{D U}{\alpha_v}$ and  $Pe_l=\frac{D U}{\alpha_l}$  are the vapor and liquid Peclet numbers respectively (representing the ratio of convection by thermal diffusion). We will solve equation \ref{eqNDdeltaWB1} by Runge Kutta $4^{th}$ order method, for the initial conditions obtained by imposing $\left.\frac{d\delta}{d\theta}\right|_{\theta =0}=0$. The condition is justified owing to the fact that the vapor layer thickness is initially very small, and grows along the sphere surface due to the addition of vapor because of boiling. Therefore, at $\theta=0$ this increase in vapor layer is negligible.\\

\subsubsection{\textbf{Initial condition}}
We will now use $\left.\frac{d\delta}{d\theta}\right|_{\theta =0}=0$ or $\left.\frac{d\left(\frac{\delta}{D}\right)}{d\theta}\right|_{\theta =0}=0$ in equation \ref{eqNDdeltaWB1} to get the initial conditions.

\begin{equation}
\begin{aligned}
\Bigg( \frac{1}{1+ \frac{3\rho_l}{2\rho_v}Re_v (\frac{\delta}{D})^2 \cos\theta + \frac{1}{3}(\frac{\delta}{D})^2 \frac{G_r}{Re_v}} \left(\frac{2 J_v}{3 Pe_v\sin\theta(\frac{\delta}{D})}+\frac{2 q_r}{3 \rho_v U h'_{fg} \sin \theta} -\right. \\ \\ \left. 2\left(\frac{\delta}{D}\right)\cot\theta-\frac{1}{2} \frac{\rho_l}{\rho_v}Re_v \left(\frac{\delta}{D}\right)^3(\frac{3\cos^2\theta -1}{\sin\theta})-
\frac{2}{9}\frac{Gr}{Re_v}\left(\frac{\delta}{D}\right)^3 \cot\theta - \right. \\ \frac{2 \frac{\rho_l}{\rho_v} J_l \sin\theta}{3\left( \frac{\pi Pe_l}{3}\left(\frac{2}{3}-\cos\theta + \frac{\cos^3\theta}{3}  \right) \right)^\frac{1}{2} } \Bigg) \Bigg)  \Bigg|_{\theta=0}=0
\label{eqinitial1}
\end{aligned}
\end{equation}

$\implies$

\begin{equation}
\begin{aligned}
\left(\frac{2 J_v}{3 Pe_v\sin\theta(\frac{\delta}{D})}+\frac{2 q_r}{3 \rho_v U h'_{fg} \sin \theta} -\right. \left. 2\left(\frac{\delta}{D}\right)\cot\theta-\frac{1}{2} \frac{\rho_l}{\rho_v}Re_v \left(\frac{\delta}{D}\right)^3(\frac{3\cos^2\theta -1}{\sin\theta})-
\frac{2}{9}\frac{Gr}{Re_v}\left(\frac{\delta}{D}\right)^3 \cot\theta - \right. \\ \frac{2 \frac{\rho_l}{\rho_v} J_l \sin\theta}{3\left( \frac{\pi Pe_l}{3}\left(\frac{2}{3}-\cos\theta + \frac{\cos^3\theta}{3}  \right) \right)^\frac{1}{2} } \Bigg)  \Bigg|_{\theta=0}=0
\label{eqinitial2}
\end{aligned}
\end{equation}

Multiplying above equation by $\frac{\delta}{D}\sin\theta$ we get

\begin{equation}
\begin{aligned}
\Bigg( \frac{2 J_v}{3 Pe_v} + \frac{2 q_r \frac{\delta}{D}}{3 \rho_v U h'_{fg}}
-2\cos\theta\left(\frac{\delta}{D}\right) ^2-\frac{1}{2} \frac{\rho_l}{\rho_v}Re_v(3\cos^2\theta -1)\left(\frac{\delta}{D}\right)^4- \frac{2}{9}\frac{Gr}{Re_v}\cos\theta \left(\frac{\delta}{D}\right)^4 -\\ \frac{2 \frac{\rho_l}{\rho_v} J_l \sin ^2\theta \frac{\delta}{D}}{3\left( \frac{\pi Pe_l}{3}\left(\frac{2}{3}-\cos\theta + \frac{\cos^3\theta}{3}  \right)    \right)^\frac{1}{2} } \Bigg) \Bigg|_{\theta=0}=0
\label{eqinitial3}
\end{aligned}
\end{equation}

$\implies$

\begin{equation}
\frac{2 J_v}{3 Pe_v}+\frac{2 q_r \frac{\delta}{D}}{3 \rho_v U h'_{fg}} - 2 \left(\frac{\delta}{D}\right) ^2-\frac{\rho_l}{\rho_v}Re_v\left(\frac{\delta}{D}\right)^4 -\frac{2}{9}\frac{Gr}{Re_v}\left(\frac{\delta}{D}\right)^4
-\left. \frac{2 \frac{\rho_l}{\rho_v} J_l \sin ^2\theta \frac{\delta}{D}}{3\left( \frac{\pi Pe_l}{3}\left(\frac{2}{3}-\cos\theta + \frac{\cos^3\theta}{3}  \right) \right)^\frac{1}{2} }\right|_{\theta=0} =0
\label{eqinitial4}
\end{equation}

The last term in equation \ref{eqinitial4} is solved separately as follows (the steps to solve last term in equation \ref{eqinitial4} is given in the appendix),

\begin{equation}
     \lim_{\theta\to 0} \left. \frac{2 \frac{\rho_l}{\rho_v} J_l \sin ^2\theta \frac{\delta}{D}}{3\left( \frac{\pi Pe_l}{3}\left(\frac{2}{3}-\cos\theta + \frac{\cos^3\theta}{3}  \right)    \right)^\frac{1}{2} }\right|_{\theta=0}= \frac{4}{\sqrt {3 \pi Pe_l}} \frac{\rho_l}{\rho_v}J_l \frac{\delta}{D}
     \label{eqinitial5}
\end{equation}

substituting equation \ref{eqinitial5} in equation \ref{eqinitial4} we get,

\begin{equation}
\frac{2 J_v}{3 Pe_v}+\frac{2 q_r}{3 \rho_v U h'_{fg}}\left(\frac{\delta}{D}\right) - 2 \left(\frac{\delta}{D}\right) ^2-\frac{\rho_l}{\rho_v}Re_v\left(\frac{\delta}{D}\right)^4 -\frac{2}{9}\frac{Gr}{Re_v}\left(\frac{\delta}{D}\right)^4
- \frac{4}{\sqrt {3 \pi Pe_l}} \frac{\rho_l}{\rho_v}J_l \left(\frac{\delta}{D}\right)=0
\end{equation}

$\implies$

\begin{equation}
\left(\frac{\rho_l}{\rho_v}Re_v +\frac{2}{9}\frac{Gr}{Re_v}\right) \left(\frac{\delta}{D}\right)^4 + 2 \left(\frac{\delta}{D}\right) ^2+
 \left( \frac{4}{\sqrt {3 \pi Pe_l}} \frac{\rho_l}{\rho_v}J_l - \frac{2 q_r}{3 \rho_v U h'_{fg}}\right)\left(\frac{\delta}{D}\right) - \frac{2 J_v}{3 Pe_v}=0
 \label{biquad}
\end{equation}

Equation \ref{biquad} is solved and the real, non-negative values of $\frac{\delta}{D}$ are the initial condition to solve equation \ref{eqNDdeltaWB1}. Equation \ref{biquad} consists of various non-dimensional terms which are required to be evaluated from the properties of vapor and liquid evaluated at corresponding mean film temperature. The mean film temperature for vapor is $\frac{T_b+T_{sat}}{2}$ and for liquid is $\frac{T_{sat}+T_w}{2}$. Note that equation \ref{eqNDdeltaWB1} has a singularity at $\theta=0^\circ \,\, \mathrm{and} \,\, 180^\circ$. Therefore the initial condition required to solve equation \ref{eqNDdeltaWB1} by Runge-Kutta method is given at some $\theta$ near to $0^\circ$ and not exactly at $\theta = 0^\circ$.\\ 

After calculating the variation of $\delta(\theta)$ we can compute the heat transfer coefficient, and the Nusselt number. We consider the fact that the energy leaving the sphere surface has two components namely conduction across the vapor film, and radiation (see figure \ref{energybal1}). Therefore, an energy balance enables us to write:

\begin{equation}
    h_\theta(T_b - T_{sat})= \frac{k_v (T_b - T_{sat})}{\delta} + q_r,
    \label{htheta1}
\end{equation}

\begin{equation}
    h_\theta = \frac{k_v}{\delta} + \frac{q_r}{T_b - T_{sat}},
    \label{htheta2}
\end{equation}

where $h_\theta$ represents the local heat transfer coefficient. The local Nusselt number,$Nu_{\theta}$ is defined as $\frac{h_{\theta} D}{k_{v}}$, and therefore, can be written as,

\begin{equation}
    Nu_\theta = \frac{D}{\delta} + \frac{D q_r}{k_v (T_b - T_{sat})}.
    \label{eqlocalNuR}
\end{equation}

From the local Nusselt number, we can calculate the average Nusselt number using the total sphere area $4\pi R^2$ as follows: 

 \begin{equation}
  Nu = \frac{1}{2}  \int_{0}^{\theta_s} Nu_\theta\sin\theta d\theta,
  \label{eqavgNu}
\end{equation}

here $\theta_s$ is the angle at which separation takes place. The reason for using $\theta_s$ in equation \ref{eqavgNu} indicates the validation of the analytical solutions till the point of separation. We further use \ref{eqavgNu}, to calculate the averaged heat transfer coefficient:

\begin{equation}
  h= \frac{Nu D}{k_v}.
  \label{eqhfromNu}
\end{equation}

\subsubsection{\textbf{Flow Separation}} \label{secflowseparation}
We can use equation \ref{velvap}, and apply $\left.\frac{\partial{u}}{\partial{y}} \right|_{y=0} = 0$ to determine the angle of separation. This boundary condition is used because the point at which the flow separates will have a velocity profile such that the gradient of velocity with respect to normal to surface becomes zero.

\begin{equation}
 \cos\theta_s = -  \left( \frac{\frac{3U}{2\delta^2_s} + \frac{\Delta \rho g}{2\mu_v}}{\frac{9 \rho_l U^2}{8\mu_v R}}\right)  = -  \left(\frac{4\mu_v R}{3 \rho_l U \delta_s ^2} + \frac{4 R g \Delta \rho}{9 U^2 \rho_l} \right),
 \label{eqsepWB}
 \end{equation}

where $\delta_s$ is the thickness of the vapor layer at the point of separation. The vapor layer grows as we move in direction of $\theta$. As we reach a point where $\theta=\theta_s$, $\delta$ is equal to $\delta_s$. \\

Equation \ref{eqsepWB} does not predict the separation angle directly, rather it needs to be solved simultaneously with equation \ref{eqNDdeltaWB1}. We can also observe from equation \ref{eqsepWB} that $\theta_s > \frac{\pi}{2}$. The second term in equation \ref{eqsepWB} represents the effect of buoyancy and scales with $\frac{1}{U^2}$. This signifies that at low velocities this term will grow rapidly and will suppress the separation resulting in an increase in the separation angle (see section \ref{results}).\\

\section{Results} 
\label{results}

\begin{figure}
(a) \includegraphics[width=7.25cm]{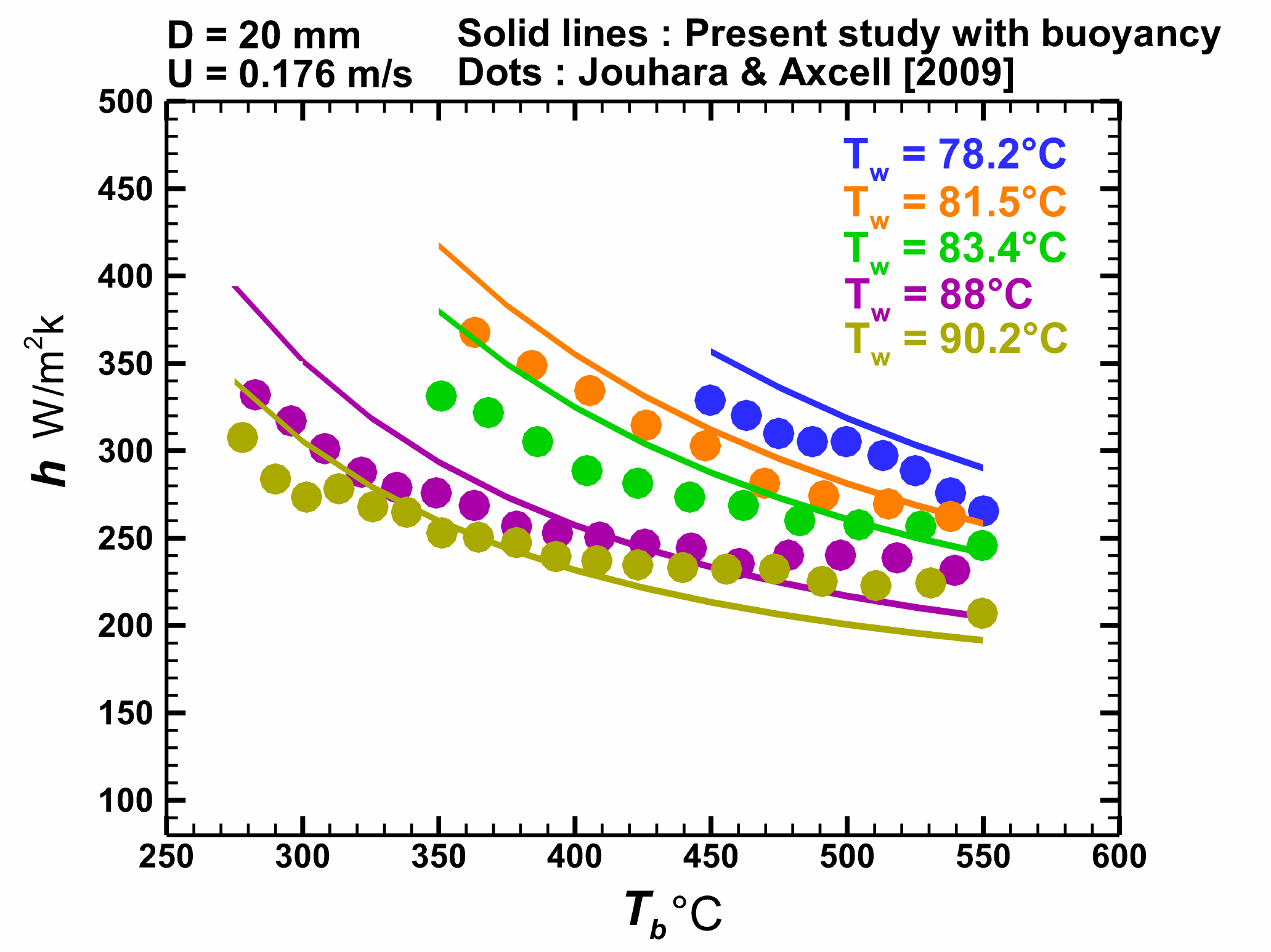}
(b) \includegraphics[width=7.25cm]{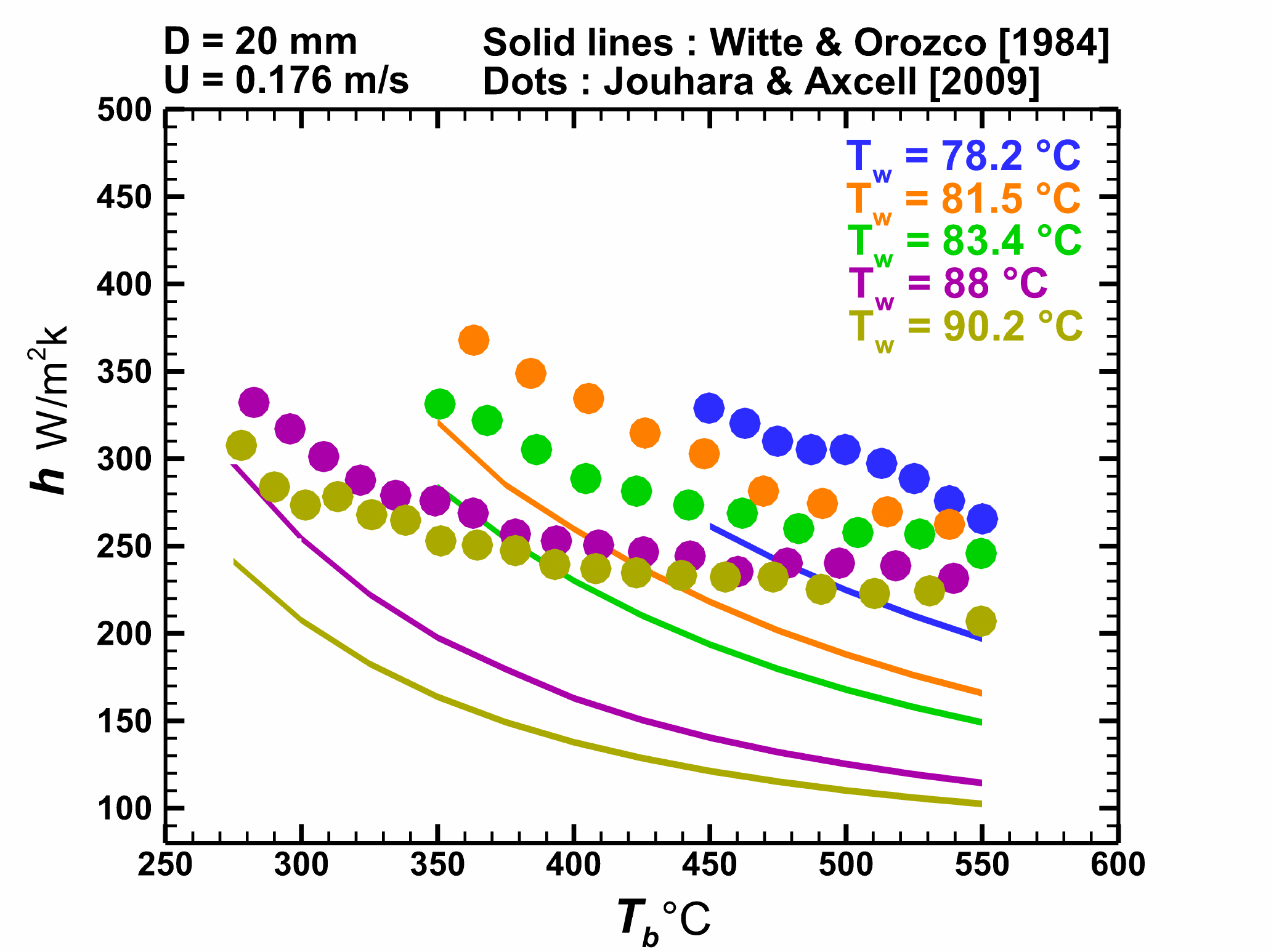}
\caption{Comparison of the heat transfer coefficient with sphere temperature between (\textit{a})  present study, and the experiment of \cite{Jouhara09}, (\textit{b}) \cite{Witte1984} and experiment by \cite{Jouhara09}. }
\label{hcomp}
\end{figure}

The comparison of the variation of heat transfer coefficient with sphere temperature between the present model, and the experiments of \cite{Jouhara09} is shown in figure \ref{hcomp} (a). We also show the corresponding comparison between the model of \cite{Witte1984}, and the experimental study of \cite{Jouhara09} in figure \ref{hcomp} (b). Our model achieves a very good agreement with the results from the experiments. The model of \cite{Witte1984} manifests a significant departure from the experimental results. The reason being the inclusion of buoyancy in our model that successfully captures the delayed separation. 
We further discuss the key role of buoyancy in delaying the separation at low velocities in the upcoming text. \\ 
\begin{figure}
(a) \includegraphics[width=7.25cm]{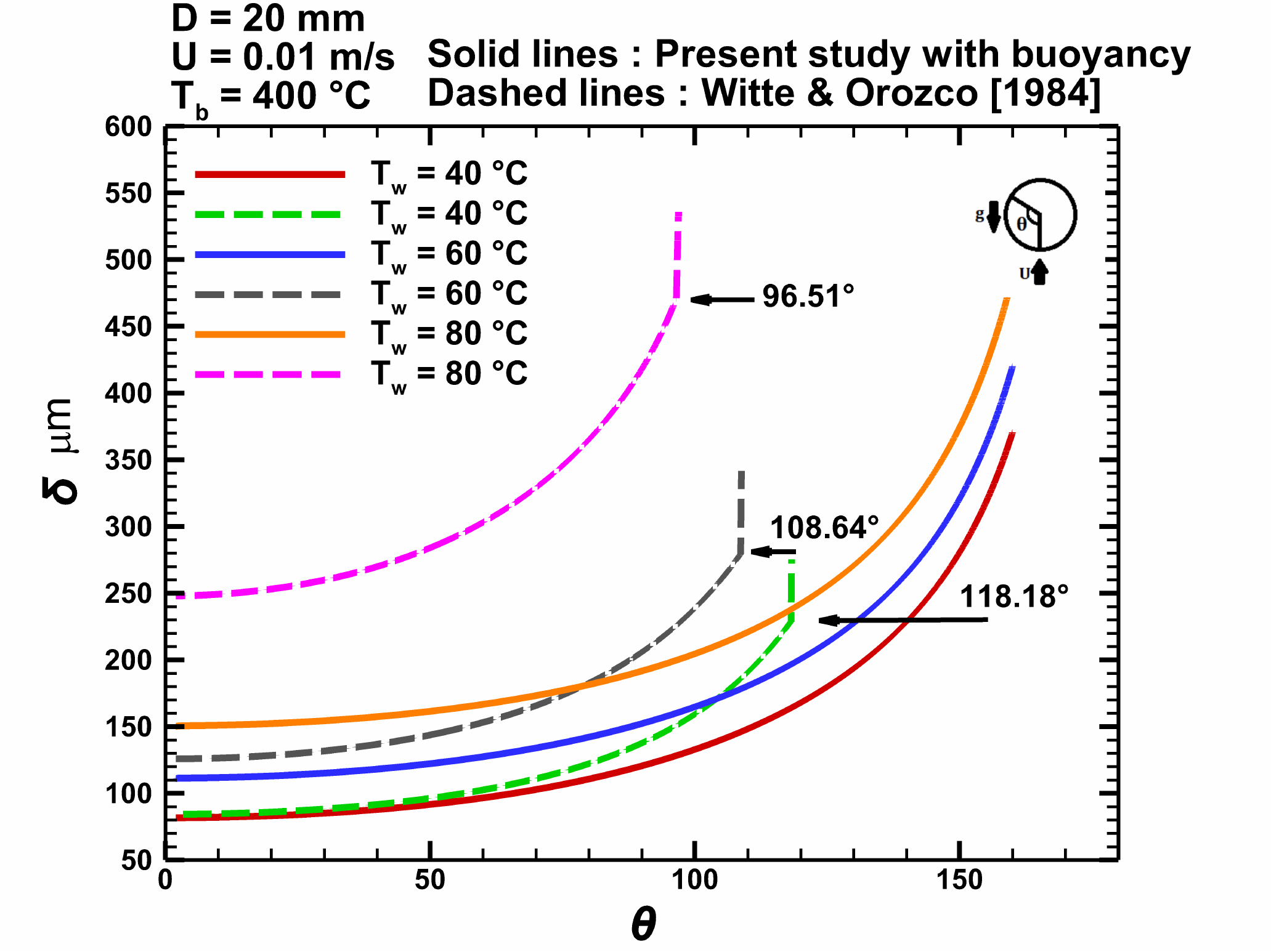}
(b)  \includegraphics[width=7.25cm]{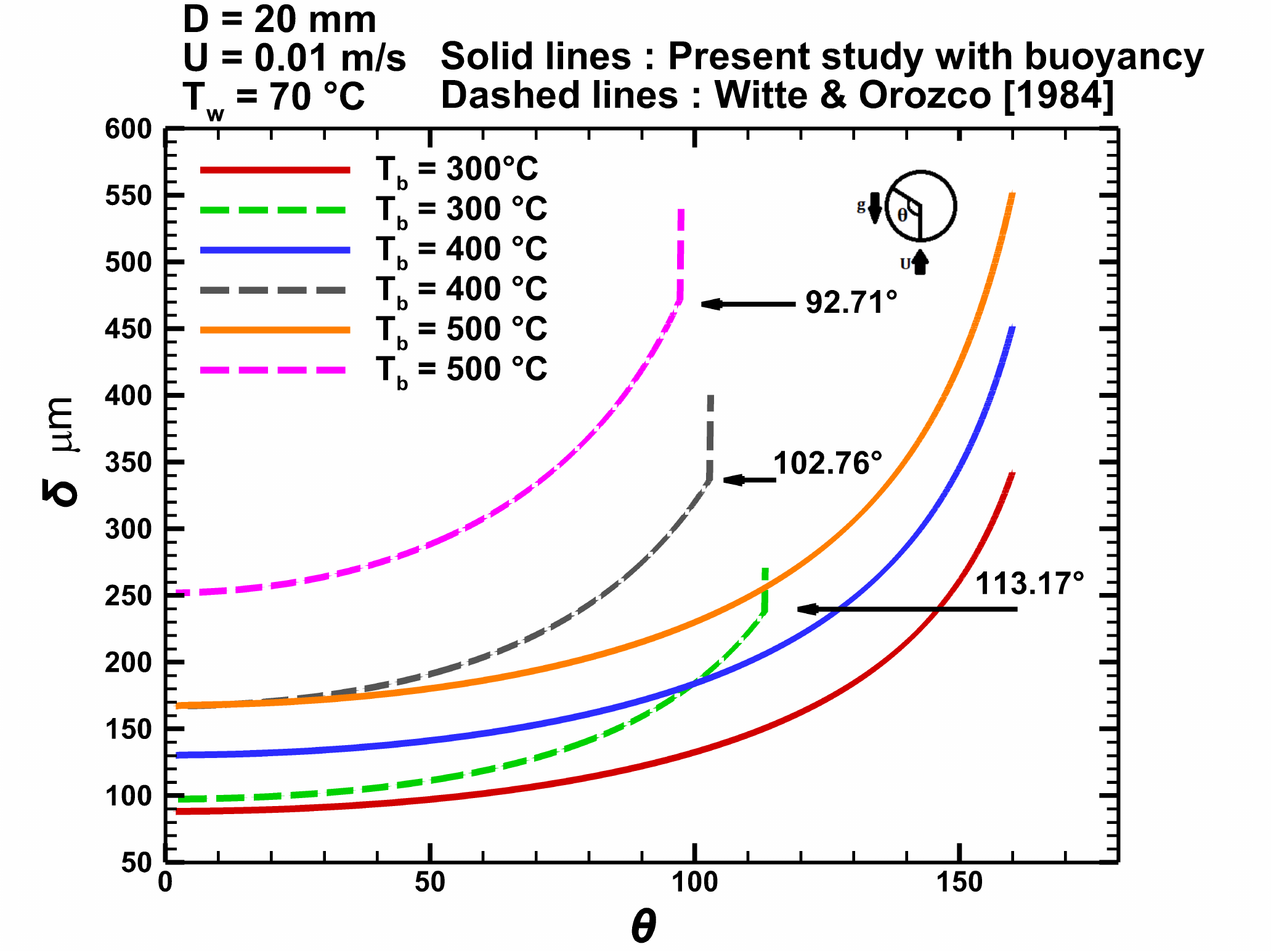}
\caption{ Variation of vapor boundary layer thickness over sphere at free stream velocity $U = 0.01$ m/s with (\textit{a}) bulk water temperature $T_w$ at  $T_b = 400^{\circ} C$, and (\textit{b}) sphere temperature $T_b$ at $T_w = 70^{\circ} C$.}
\label{VBL}
\end{figure}

The vapor boundary layer thickness $\delta$ increases with an increase in temperature of water ($T_w$), and an increase in sphere temperature ($T_b$) 
as can be observed from figures \ref{VBL} (a) and (b) respectively. With an increase in the temperature of water $T_w$, the contribution of vapor film to the net energy exchange between the sphere, and the surrounding liquid decreases. Therefore, the energy going into the bulk liquid decreases, and the amount of total energy available for vaporization of liquid increases resulting in an increase in the vapor boundary layer thickness. Similarly, with an increase in sphere temperature $T_b$, the energy available for vaporization of the liquid increases resulting in an increase in vapor boundary layer thickness.  At low velocities, the adverse pressure gradient weakens, and buoyancy (acting upwards) pushes the fluid against this weak adverse pressure gradient to delay the separation. Our model captures this separation delay for different $T_w$, and $T_b$ as manifested in figures \ref{VBL} (a) and (b). The model of \cite{Witte1984} owing to the exclusion of buoyancy predicts significantly early separation at low velocities as indicated by the rapid rise in the vapor boundary layer thickness in figures \ref{VBL} (a) and (b). \\


\begin{figure}
(a)\includegraphics[width=7.25cm]{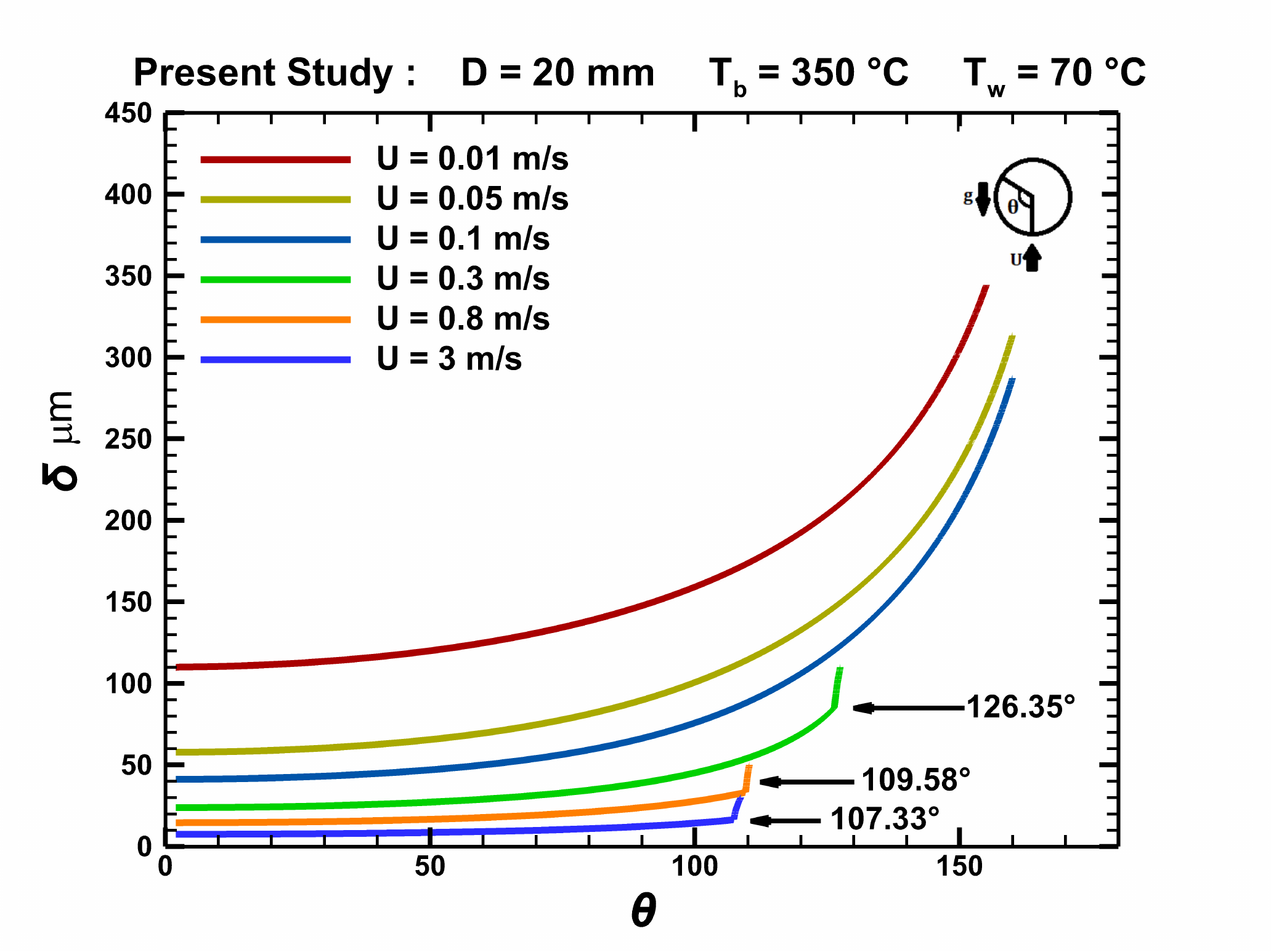}
(b)\includegraphics[width=7.25cm]{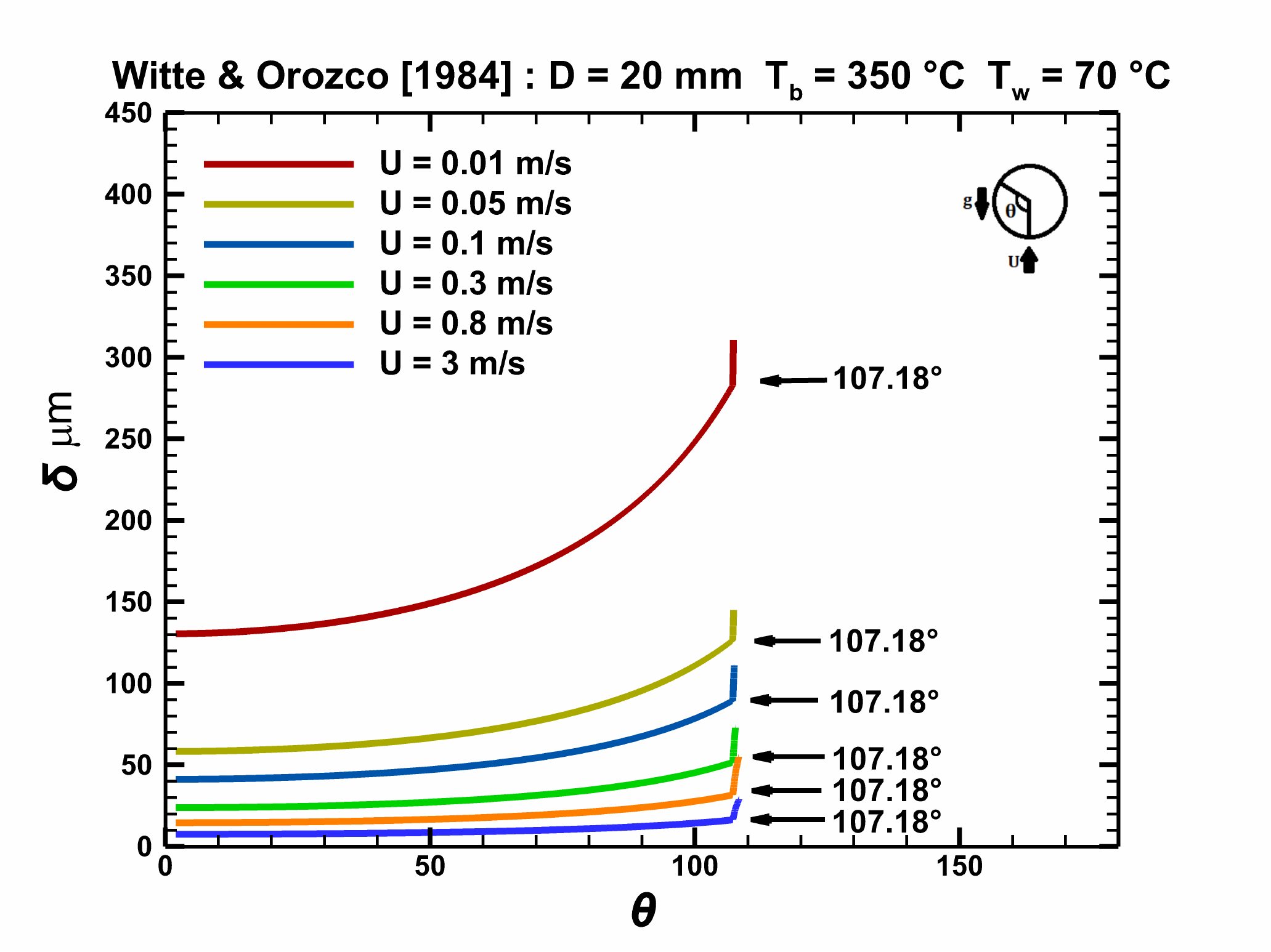}
\caption{Variation of vapor boundary layer thickness over the sphere at different velocities for given sphere and bulk water temperature obtained from (\textit{a}) present study, and (\textit{b}) the model of \cite{Witte1984}.}
\label{deltaU}
\end{figure}

We report a decrease in the vapor boundary layer thickness with an increase in free stream velocity $U$ for a given value of the sphere and the water temperature in figure \ref{deltaU}. It can be observed from figure \ref{deltaU} (a) that the separation is delayed (shown by the sudden increase in $\delta$) with decreasing velocity. When the velocity becomes sufficiently low there is no separation which is similar to the observations of \cite{bromley1953heat}. In comparison to our model, the model of \cite{Witte1984} does not show any variation of separation angle with velocity (see figure \ref{deltaU} (b)). 
To understand the reason for the separation even at low velocities we analyze the expression of the angle of separation ($\cos\theta_s = - \left(\frac{4\mu_v R}{3 \rho_l U \delta_s ^2} \right)$) from the model of \cite{Witte1984}.  
At a given $T_w$ and $T_b$ the product of $U$ and $\delta_s^2$ remains constant as shown in  table \ref{table1}. 
Therefore, the denominator in the expression of $cos\theta_s$ will remain a constant and, therefore the separation angle will remain the same with velocity.\\

\begin{table}
   \begin{center}
   \def~{\hphantom{0}}
    \begin{tabular}{ccccc}
    $U$ (m/s)& $\delta_s$ ($\mu m$)& $\theta_s$ & $U*\delta_s ^2*10^{12}$\\
    \hline
   \hline
    3&16.34&107.18$^\circ$&801.1\\
    0.8&31.64&107.18$^\circ$&801.1\\
    0.3&51.68&107.18$^\circ$&801.1\\
    0.1&89.51&107.18$^\circ$&801.1\\
    0.05&126.58&107.1$^\circ$&801.1\\
    0.01&283.03&107.18$^\circ$&801.1\\
    \hline
    \end{tabular}
    \caption{Model of \cite{Witte1984} at $T_w$ = 70$^\circ C$, $T_b$ = 350$^\circ C$, D = 20 mm }
    \label{table1}
    \end{center}
\end{table}
\begin{table}
   \begin{center}
   \def~{\hphantom{0}}
   \begin{tabular}{cccccc} 
   $U$ (m/s) & $\delta_s$ ($\mu m$)&  First term & Second term & First term + Second term & $\theta_s$ \\
    \hline
    \hline
    3 & 16.41 & 0.2931 & 0.0048 & 0.2979 & 107.33$^\circ$\\
    0.8 & 33.28 & 0.2671 & 0.0681 & 0.3352 & 109.58$^\circ$ \\
    0.5 & 45.91 & 0.2246 & 0.1744 & 0.3990 & 113.51$^\circ$\\
    0.3 & 85.26 & 0.1085 & 0.4844 & 0.5930 & 126.35$^\circ$\\
    0.1 & & & No Separation & &  \\
   \hline
   \end{tabular}
   \caption{Present model at $T_w$ = 70$^\circ C$, $T_b$ = 350$^\circ C$, D=20 mm}
   \label{table2}
   \end{center}
\end{table}

Our expression for $\theta_s$ is composed of two terms, $\frac{4\mu_v R}{3 \rho_l U \delta_s ^2}$ and $\frac{4 R g \Delta \rho}{9 U^2 \rho_l}$. The second term represents the influence of buoyancy and scales with $\frac{1}{U^2}$. Therefore, decreasing the velocity, $U$, increases the second term. Table \ref{table2} demonstrates the variation of the first and second term with velocity for a given $T_b$ and $T_w$. We can observe that as we decrease the velocity, the value of $\delta_s$ increases, the first term decreases, and the second increases. 
It can also be observed that the second term is negligible at high values of $U$, and doesn't contribute much in delaying the separation at high velocity. However, at low velocities, the contribution of buoyancy (second term) becomes significant resulting in separation delay. \\
 
\begin{figure}
\centering
\begin{subfigure}[h]{0.7\textwidth}
      \centering
      \includegraphics[width=\textwidth]{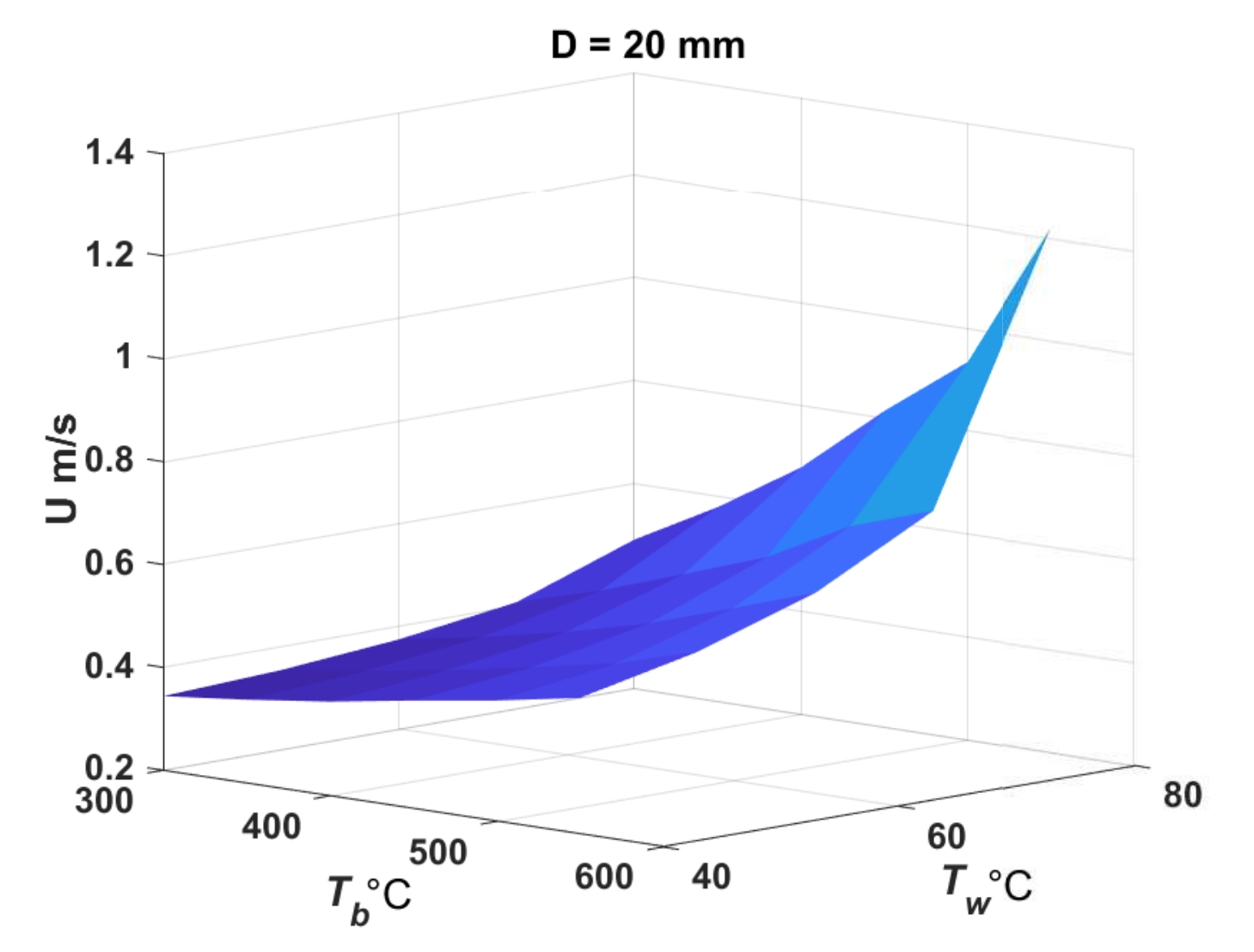}
      \caption{Surface plot of velocity, at which the first and the second terms in the expression for separation angle become equal, at different sphere and bulk water temperature obtained from the present study.}  
      \label{sepsurf1}
\end{subfigure}
\begin{subfigure}[h]{0.7\textwidth}
      \centering
      \includegraphics[width=\textwidth]{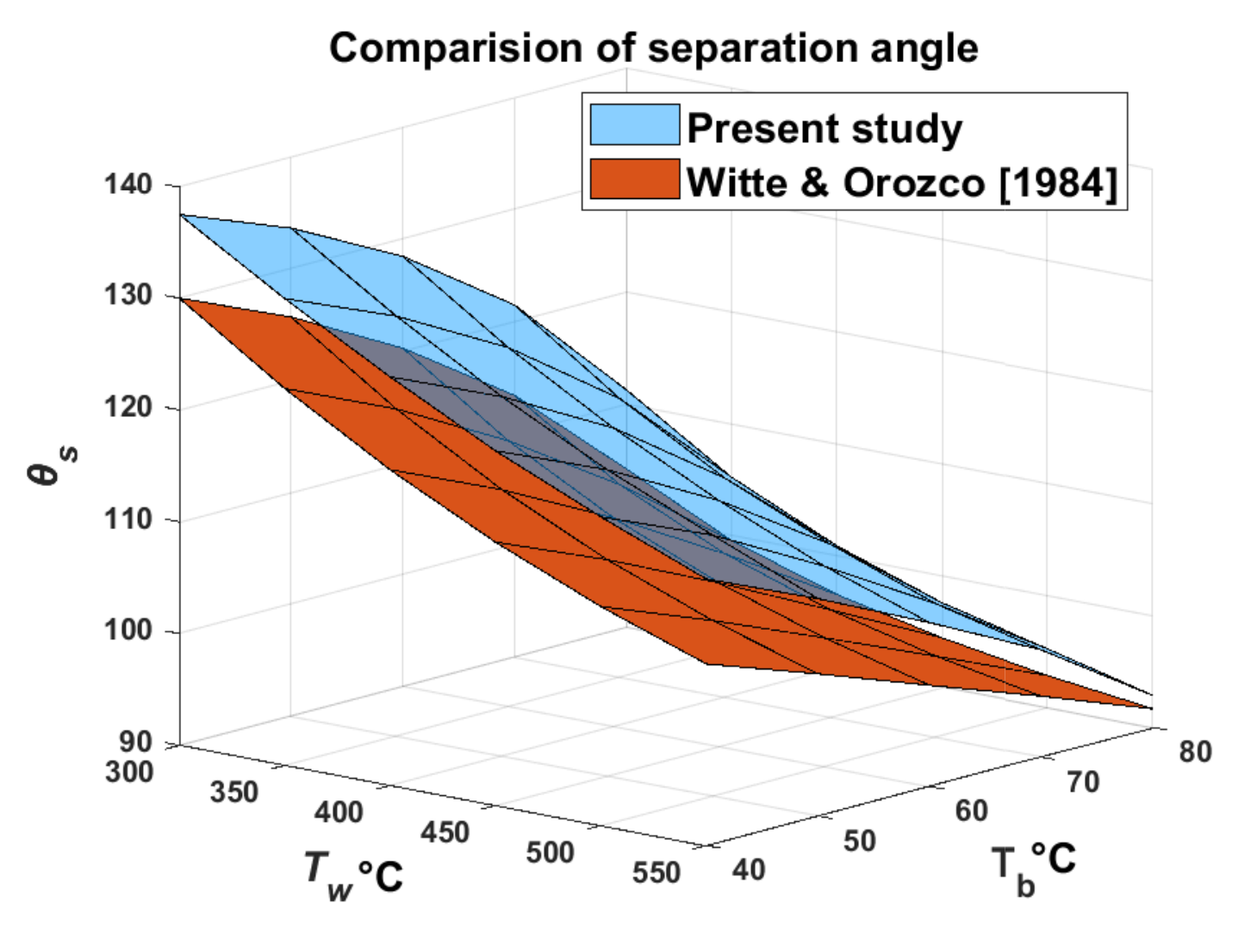}
      \caption{Comparison of the surface plot of separation angle for the corresponding parameters of figure \ref{sepsurf1} obtained from present study and model of \cite{Witte1984}. \protect} 
      \label{sepsurf2}
\end{subfigure}
\caption{}
\end{figure}

\begin{table}
   \begin{center}
   \def~{\hphantom{0}}
    \begin{tabular}{cccccccc} 
    \hline
    $T_w\, ^\circ$C & $T_b\, ^\circ$C & $U$ (m/s) & $\delta_s$ ($\mu m$)& First term & Second term & First term + Second term & $\theta_s$ \\
    \hline
    \hline
    40 & 300 & 0.344 & 41.72 & 0.3684 & 0.3684 & 0.7368 & 137.46$^\circ$ \\
    70 & 350 & 0.455 & 49.72 & 0.2105 & 0.2105 & 0.4210 & 114.89$^\circ$ \\
    80 & 400 & 0.68  & 62.65 & 0.0943 & 0.0943 & 0.1886 & 100.87$^\circ$ \\        
    80 & 450 & 0.82  & 70.46 & 0.0657 & 0.0657 & 0.1314 & 97.55$^\circ$ \\  
    80 & 550 & 1.22  & 95.01 & 0.0280 & 0.0280 & 0.0575 & 93.30$^\circ$ \\
    \hline
    \end{tabular}
    \caption{A representative dataset for figure \ref{sepsurf1}}
    \label{table3}
    \end{center}
\end{table}

We create a three-dimensional surface plot (see figure \ref{sepsurf1}) of the velocity, at which the first and the second terms in the expression for separation angle becomes equal, at different sphere, and water temperature.
We present a data set comprising of the values of the first and the second terms in the expression for separation angle in table \ref{table3}. Comparative three-dimensional surface plots (figure \ref{sepsurf2}) of the variation of the separation angle with respect to the sphere, and the water temperature are also generated by using the corresponding parameters of figure \ref{sepsurf1} for the present model, and the model of \cite{Witte1984}. The region in figure \ref{sepsurf2}, where the surface generated by the present model is away from the surface generated by the model of \cite{Witte1984} corresponds to a region of low velocities in figure \ref{sepsurf1}. Similarly, the region in figure \ref{sepsurf2}, where the two surfaces are closer to each other, corresponds to a region of high velocities in figure \ref{sepsurf1}. Therefore, even when the first and the second terms in the expression of separation angle are equal the buoyancy may not be significant as can be seen at high velocities where the difference in the separation angle for the present model with the model of \cite{Witte1984} diminishes. Hence, we can conclude that at a particular sphere, and bulk water temperature, the influence of buoyancy is significant at lower velocities where the difference in the separation angle for present model and model of \cite{Witte1984} is significant. Clearly the model of \cite{Witte1984} under predicts the separation angle at all velocities owing to the exclusion of buoyancy in their analysis.\\

\begin{figure}
(a) \includegraphics[width=7.25cm]{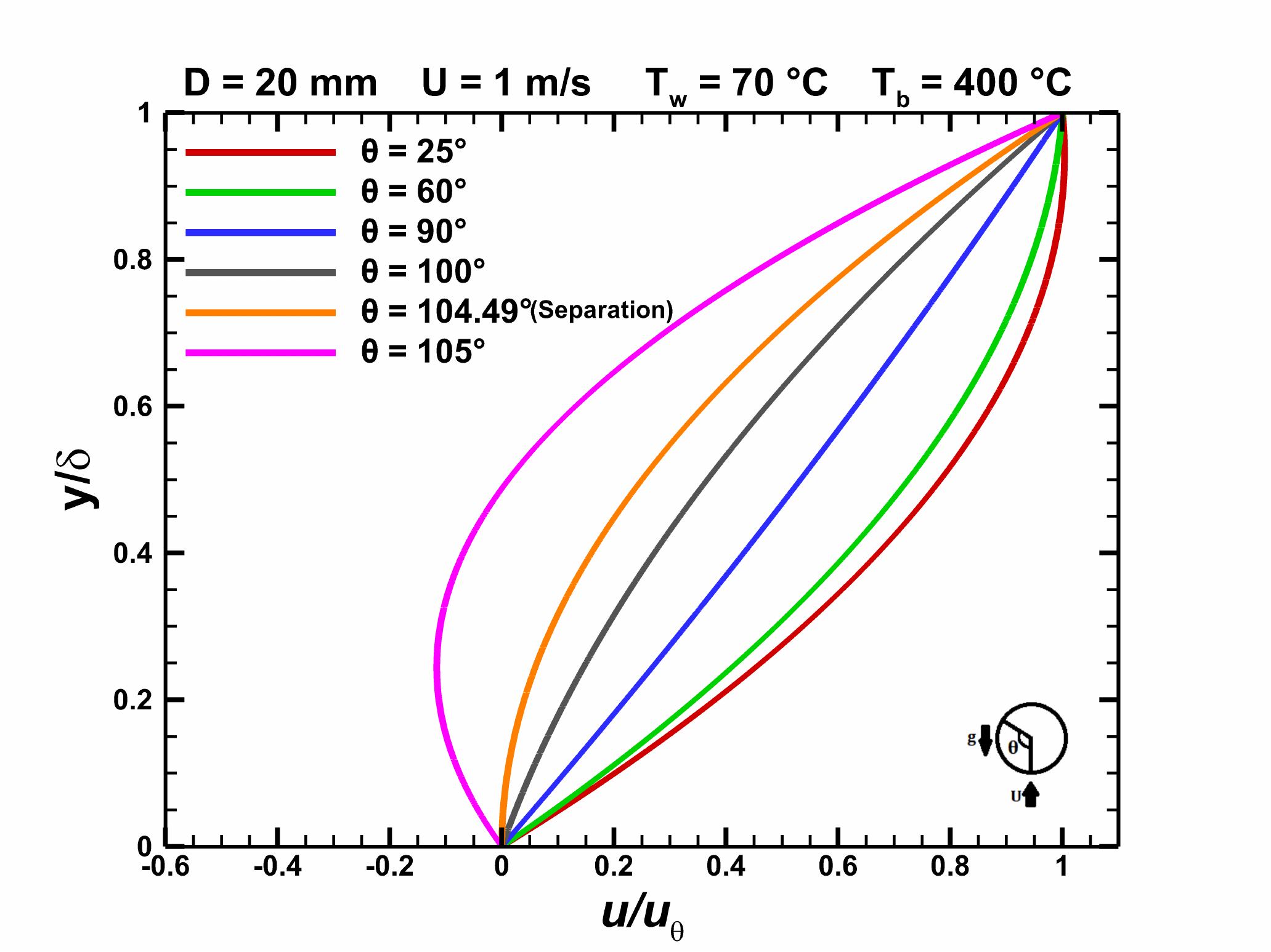}
(b) \includegraphics[width=7.25cm]{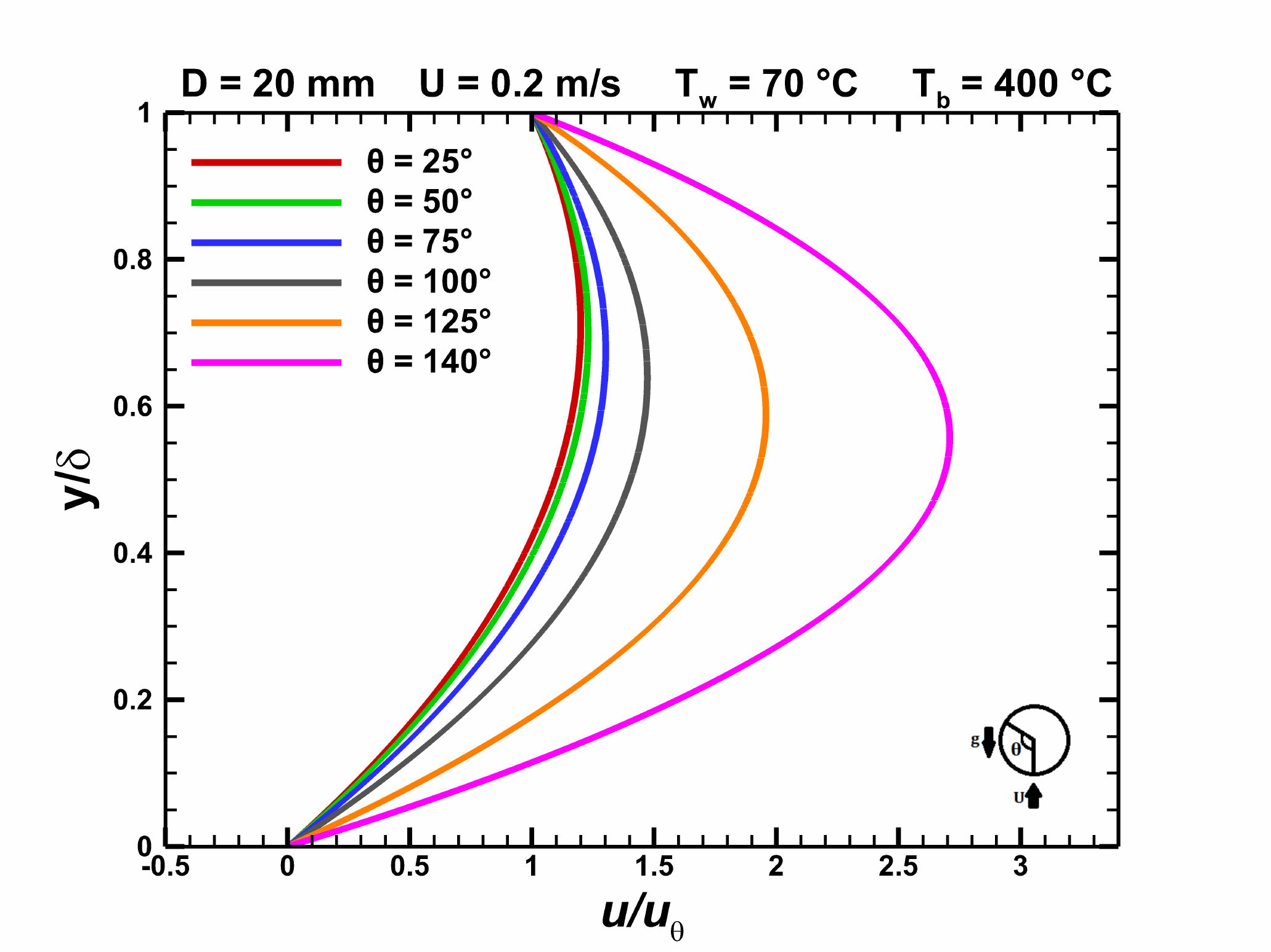}
\caption{Non-dimensional velocity profile (refer equation \ref{equinliq} for the expression of $u_\theta$) at (\textit{a}) $U = 1$ m/s, $T_w = 70 ^\circ$C, $T_b = 400 ^\circ$C, and (\textit{b}) $U = 0.2$ m/s, $T_w = 70 ^\circ$C, $T_b = 400 ^\circ$C }  
\label{NDvel}
\end{figure}

\begin{figure}
(a) \includegraphics[width=7.25cm]{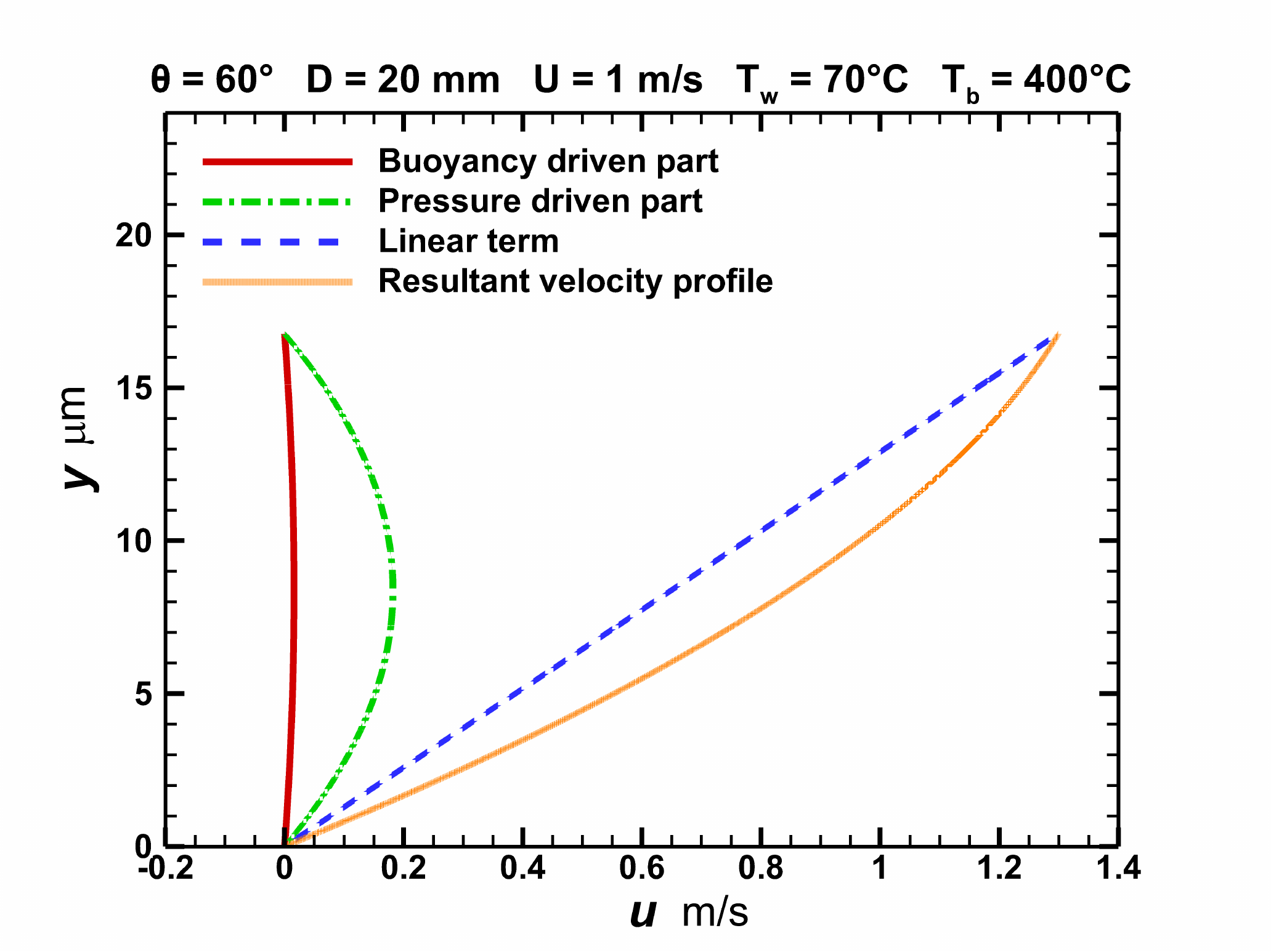}
(b) \includegraphics[width=7.25cm]{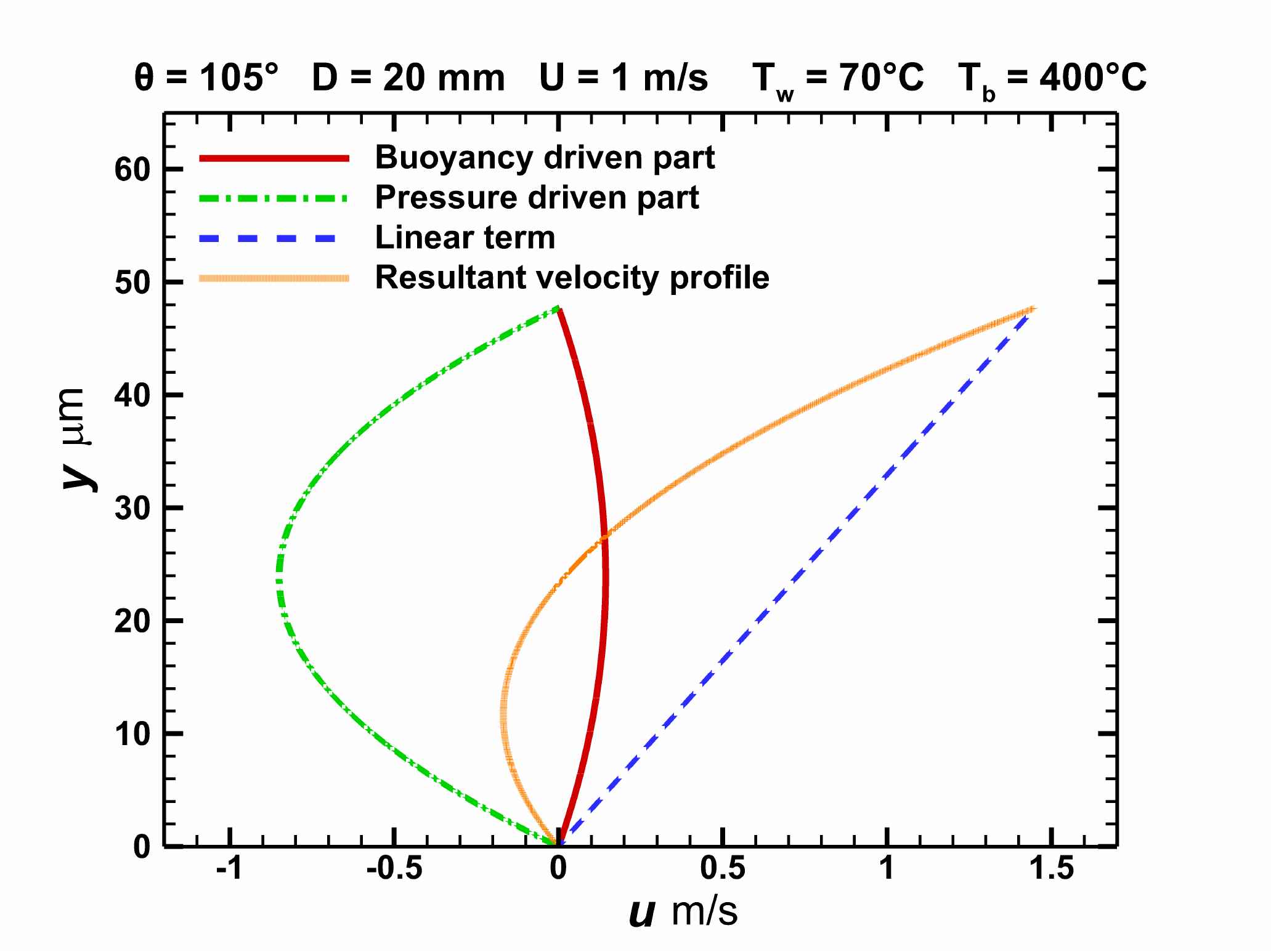}
(c) \includegraphics[width=7.25cm]{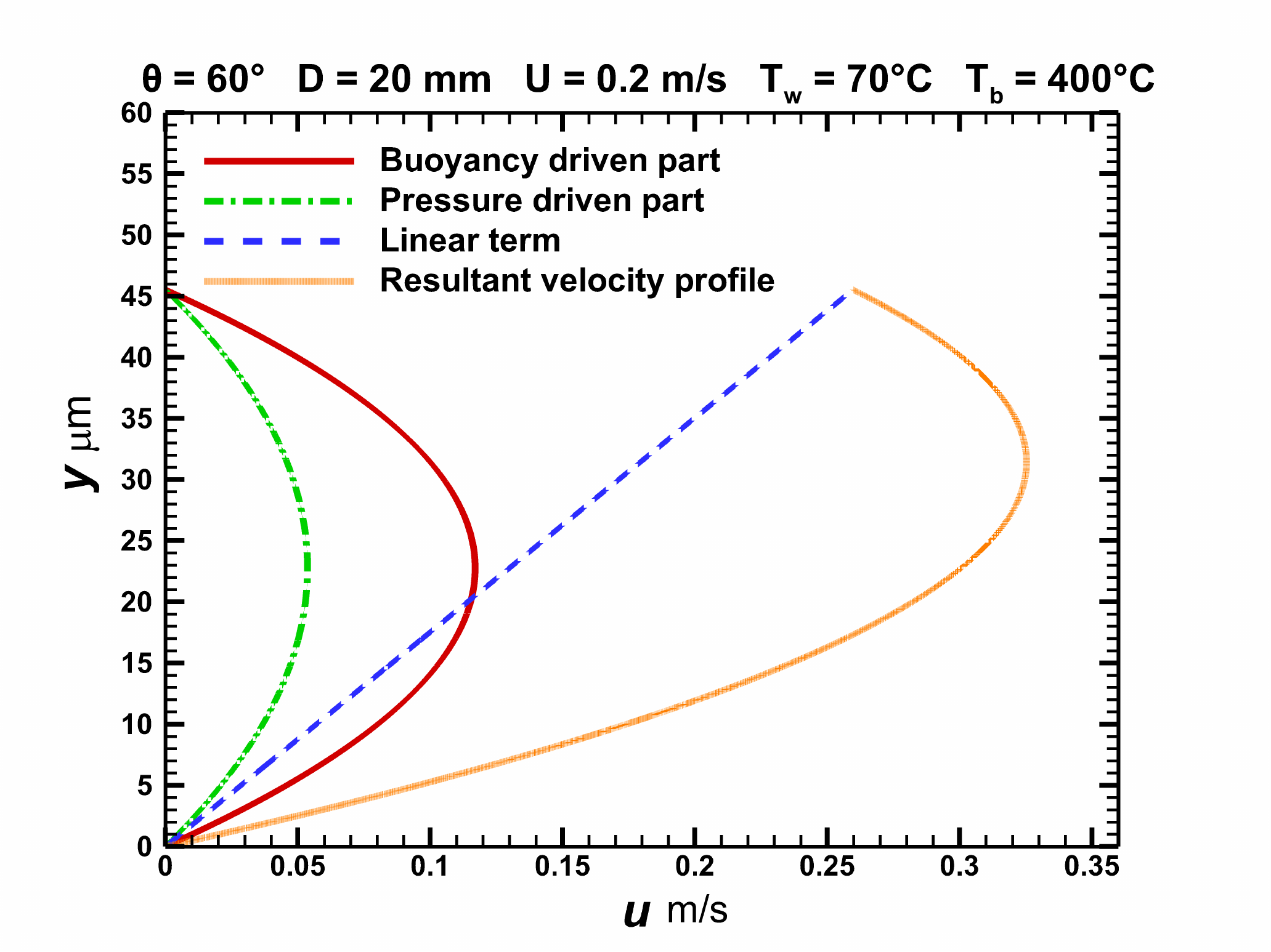}
(d) \includegraphics[width=7.25cm]{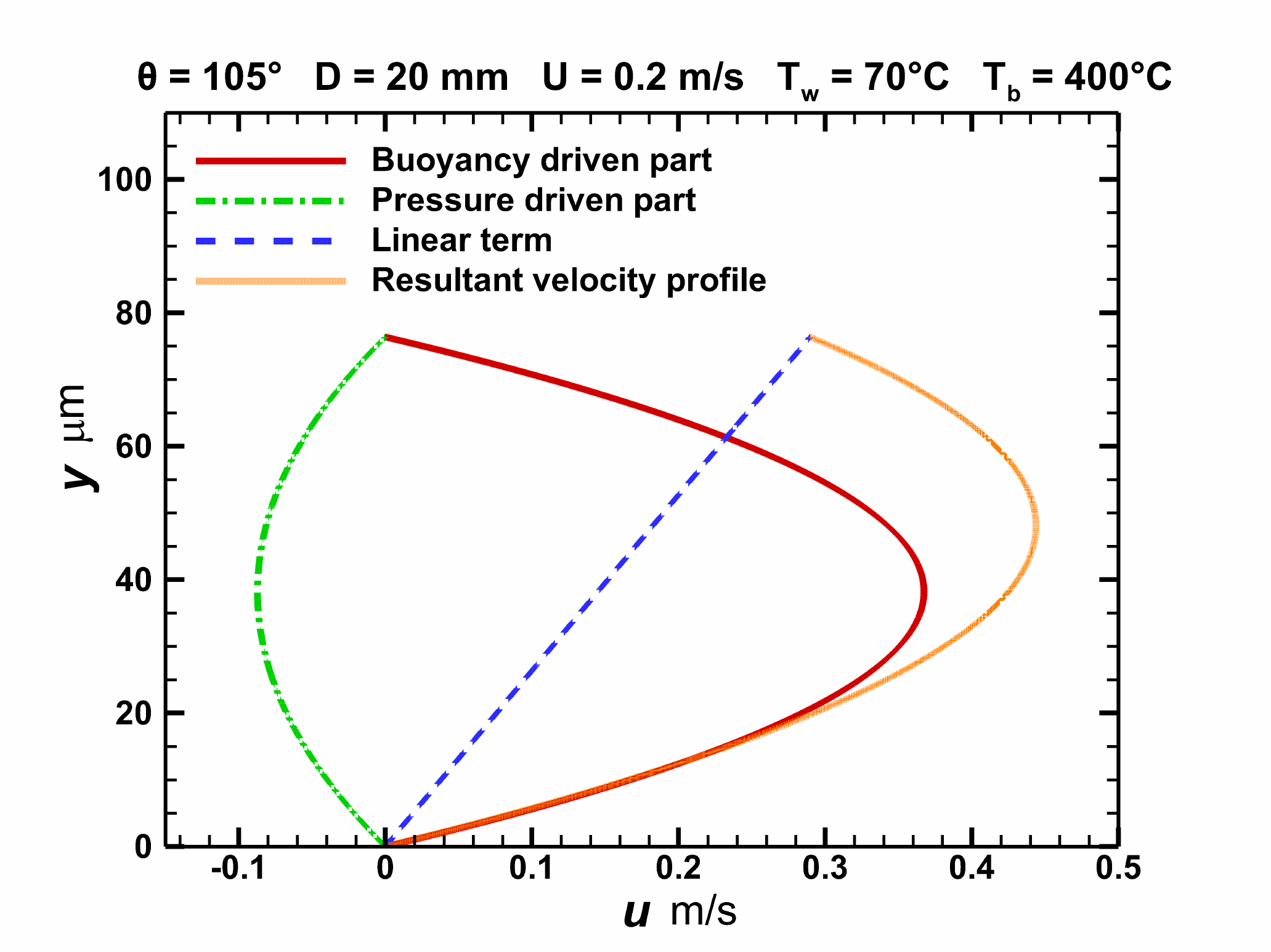}
\caption{Dimensional velocity profile (refer equation \ref{velvap}) at different values of $\theta$ for $U=1$ m/s and $U=0.2$ m/s }
\label{combinedUprofile}
\end{figure}

According to equation \ref{momeqvap3} the pressure gradient is favorable in the bottom half of the sphere ($\theta<90^\circ$) whereas it is adverse in the top half ($\theta>90^\circ$) of the sphere. Buoyancy favors the flow of the vapor in both the lower and the upper halves. When the velocity is high the adverse pressure gradient in the top half is also large, and even though buoyancy supports the vapor flow, the flow may separate. As the velocity decreases, the adverse pressure gradient in the top half of the sphere decreases, and buoyancy dominates the vapor flow resulting in delayed or no separation. Figures \ref{NDvel} (a) and (b) represent the non-dimensional velocity profiles at different angles for high, and low velocities respectively. At high velocity (figure \ref{NDvel} (a)) it can be observed that separation takes place in the top half of the sphere but at low velocity (figure \ref{NDvel}(b)) we do not observe any separation. To further access the role of buoyancy in suppressing the separation we plot the three components in equation \ref{velvap} in figures \ref{combinedUprofile} for two different velocities at two different angles. We can observe that for $U = 1$ m/s and $0.2$ m/s at $\theta = 60^\circ$, which represents a location at the bottom half of the sphere, both the pressure and the buoyancy are assisting the flow of vapor (see figures \ref{combinedUprofile} (a) and (c)). However, at $\theta = 105^\circ$, which represents a location at the upper half of the sphere, the pressure gradient is adverse and it competes with buoyancy to get the flow separated for $U = 1$ m/s and delay the separation for $U = 0.2$ m/s (figures \ref{combinedUprofile} (b) and (d) respectively).\\

\section{Conclusion}
\label{summary}
A theoretical investigation is performed to understand the influence of buoyancy on the heat transfer characteristics and  boundary layer separation behavior due to film boiling from a slowly moving heated sphere. The novelty of this study lies in the inclusion of the buoyancy in the governing equation unprecedented to the previous theoretical investigations for a spherical body. In the present analytical model the momentum, and the energy equations are solved in the vapor phase to obtain the velocity, and the temperature distribution in terms of the vapor layer thickness. We apply an energy balance at the vapor-liquid interface to determine the vapor layer thickness. The flow of liquid around the sphere is considered to be governed
by potential theory, and the energy equation in liquid is then solved for the known velocity
distribution.\\

We find that the film boiling heat transfer coefficient decreases with an increase in sphere, and bulk water temperature owing to a subsequent increase in the vapor layer thickness. This behavior of the heat transfer coefficient resembles closely to the experimental results reported by \cite{Jouhara09}. We also found that buoyancy plays a very significant role at low velocities in delaying the separation and allows the heat transfer calculation from a larger area. 
We have included buoyancy in the expression of the vapor velocity that resulted in capturing the delayed separation phenomenon similar to the observations of by \cite{bromley1953heat} and \cite{kobayasi1965film}. 
We further analyzed the dependence of the flow separation behavior on the relative magnitudes of the pressure gradient, and buoyancy. At high velocity the pressure gradient overshadows the  buoyancy effects, and the flow separates. However, at sufficiently low velocities buoyancy drives the flow against the adverse pressure gradient, and separation is not observed. Therefore, it can be concluded that the inclusion of buoyancy is imperative for capturing the  correct variation of the heat transfer characteristics, and the boundary layer separation phenomenon at low velocities. \\

\bibliographystyle{jfm}
\bibliography{jfm}

\begin{thebibliography}{17}
\expandafter\ifx\csname natexlab\endcsname\relax\def\natexlab#1{#1}\fi
\def\au#1{#1} \def\ed#1{#1} \def\yr#1{#1}\def\at#1{#1}\def\jt#1{\textit{#1}}
  \def\bt#1{#1}\def\bvol#1{\textbf{#1}} \def\vol#1{#1} \def\pg#1{#1}
  \def\publ#1{#1}\def\arxiv#1{#1}\def\org#1{#1}\def\st#1{\textit{#1}}

\bibitem[Bradfield(1966)]{Bradfield66}
{\sc \au{Bradfield, W.~S.}} \yr{1966}  \at{Liquid-solid contact in stable film
  boiling.}  \jt{Ind. and Eng. Chemn Fundamentals}  \bvol{5},  \pg{201--204}.

\bibitem[Bradfield(1967)]{bradfield1967effect}
{\sc \au{Bradfield, W.~S.}} \yr{1967}  \at{{On the Effect of Subcooling on Wall
  Superheat in Pool Boiling}}.  \jt{Journal of Heat Transfer}  \bvol{89}~(3),
  \pg{269--270}.

\bibitem[Bromley {\em et~al.\/}(1953)Bromley, LeRoy \&
  Robbers]{bromley1953heat}
{\sc \au{Bromley, LeRoy~A}, \au{LeRoy, Norman~R} \& \au{Robbers, James~A}}
  \yr{1953}  \at{Heat transfer in forced convection film boiling}.
  \jt{Industrial \& Engineering Chemistry}  \bvol{45}~(12),  \pg{2639--2646}.

\bibitem[Burns(1989)]{Burns89}
{\sc \au{Burns, R.~A.}} \yr{1989}  \at{Heat transfer studies with application
  to nuclear reactors}. PhD thesis, Thesis, University of Manchester.

\bibitem[Dhir \& Purohit(1978)]{Dhir78}
{\sc \au{Dhir, V.K.} \& \au{Purohit, G.P.}} \yr{1978}  \at{Subcooled
  film-boiling heat transfer from spheres}.  \jt{Nuclear Engineering and
  Design}  \bvol{47}~(1),  \pg{49--66}.

\bibitem[Hesson \& Witte(1966)]{hesson1966comment}
{\sc \au{Hesson, JC} \& \au{Witte, LC}} \yr{1966}  \at{Comment on “film
  boiling heat transfer around a sphere in forced convection” by k.
  kobayasi}.  \jt{Journal of Nuclear Science and Technology}  \bvol{3}~(10),
  \pg{448--449}.

\bibitem[Jouhara \& P.Axcell(2009)]{Jouhara09}
{\sc \au{Jouhara, H.} \& \au{P.Axcell, B.}} \yr{2009}  \at{Film boiling heat
  transfer and vapour film collapse on spheres, cylinders and plane surfaces.}
  \jt{Nuclear Engineering and Design.}  \bvol{239},  \pg{1885--1900}.

\bibitem[Kobayasi \& Kiyosi(1965)]{kobayasi1965film}
{\sc \au{Kobayasi} \& \au{Kiyosi}} \yr{1965}  \at{Film boiling heat transfer
  around a sphere in forced convection}.  \jt{Journal of Nuclear Science and
  Technology}  \bvol{2}~(2),  \pg{62--67}.

\bibitem[Kutateladze(1959)]{Kutateladze59}
{\sc \au{Kutateladze, S.~S.}} \yr{1959}  \at{{Liquid-Metal Heat Transfer
  Media}}.  \bt{In {\em In Interactive dynamics of convection and
  solidification\/}},  \pg{pp. 77--81}.  \publ{Springer US}.

\bibitem[Motte \& Bromley(1957)]{motte1957film}
{\sc \au{Motte, Eugene~I} \& \au{Bromley, Leroy~A}} \yr{1957}  \at{Film boiling
  of flowing subcooled liquids}.  \jt{Industrial \& Engineering Chemistry}
  \bvol{49}~(11),  \pg{1921--1928}.

\bibitem[Sideman(1966)]{Sideman66}
{\sc \au{Sideman, S.}} \yr{1966}  \at{The equivalence of the penetration theory
  and potential flow theories.}  \jt{Ind. and Eng. Chemn Fundamentals}
  \bvol{58},  \pg{54--58}.

\bibitem[Vakarelski {\em et~al.\/}(2011)Vakarelski, Marston, Chan \&
  Thoroddsen]{vakarelski2011drag}
{\sc \au{Vakarelski, Ivan~U}, \au{Marston, Jeremy~O}, \au{Chan, Derek~YC} \&
  \au{Thoroddsen, Sigurdur~T}} \yr{2011}  \at{Drag reduction by leidenfrost
  vapor layers}.  \jt{Physical review letters}  \bvol{106}~(21),  \pg{214501}.

\bibitem[Walford(1969)]{walford1969transient}
{\sc \au{Walford, FJ}} \yr{1969}  \at{Transient heat transfer from a hot nickel
  sphere moving through water}.  \jt{International Journal of Heat and Mass
  Transfer}  \bvol{12}~(12),  \pg{1621--1625}.

\bibitem[Witte(1967)]{Witte67}
{\sc \au{Witte, L.~C.}} \yr{1967}  \at{heat transfer from a sphere to liquid
  sodium during forced convection}. PhD thesis, Oklahoma State University,
  https://digital.library.unt.edu/ark:/67531/metadc1024053/.

\bibitem[Witte(1968{\natexlab{{\em a\/}}})]{witte1968a}
{\sc \au{Witte, L.~C.}} \yr{1968{\natexlab{{\em a\/}}}}  \at{{An Experimental
  Study of Forced-Convection Heat Transfer From a Sphere to Liquid Sodium}}.
  \jt{Journal of Heat Transfer}  \bvol{90}~(1),  \pg{9--12}.

\bibitem[Witte(1968{\natexlab{{\em b\/}}})]{witte1968b}
{\sc \au{Witte, L.~C.}} \yr{1968{\natexlab{{\em b\/}}}}  \at{Film boiling from
  a sphere}.  \jt{Industrial \& Engineering Chemistry Fundamentals}
  \bvol{7}~(3),  \pg{517--518}.

\bibitem[Witte \& Orozco(1984)]{Witte1984}
{\sc \au{Witte, L.~C.} \& \au{Orozco, J.}} \yr{1984}  \at{{The Effect of Vapor
  Velocity Profile Shape on Flow Film Boiling From Submerged Bodies}}.
  \jt{Journal of Heat Transfer}  \bvol{106}~(1),  \pg{191--197}.

\end{thebibliography}

\appendix

\section*{Appendix} 
\label{Appendix}
\textbf{Steps to non-dimensionalize equation \ref{diffdelta} :}\\

Consider equation \ref{diffdelta},
\begin{equation}
\begin{aligned}
\frac{d\delta}{d\theta}= \frac{1}{ \frac{h'_{fg}\rho_v}{R} \left( \frac{3 U \sin\theta}{4} + \frac{9\rho_l U^2}{16\mu_v R}\sin\theta \cos\theta\delta^2 + \frac{\Delta \rho g \sin\theta}{4 \mu_v}\delta^2 \right)}\Bigg( \frac{k_v  \left(T_b - T_{sat}\right)}{\delta}\,+\,\\ \sigma\epsilon(T^4_b - T^4_{sat})\,-\frac{k_l\Delta T_w \sin ^2 \theta }{\sqrt{\pi M\eta}}\, -\frac{h'_{fg}\rho_v}{R}\left(\frac{3U \cos\theta \delta}{2} + \right. \\ \left. \frac{3 \rho_l U^2}{16\mu_v R}(3\cos^2\theta -1)\delta^3 + \frac{\Delta \rho g \cos\theta}{6\mu_v}\delta^3  \right) \Bigg)
\end{aligned}
\end{equation}

To non-dimensionalize the above equation we first divide the numerator and denominator by $\frac{h'_{fg}\rho_v}{R} \frac{3 U \sin\theta}{4} $. We first consider the non-dimensionalization of denominator as follows:

\begin{equation}
\frac{h'_{fg}\rho_v}{R} \left( \frac{3 U \sin\theta}{4} + \frac{9\rho_l U^2}{16\mu_v R}\sin\theta\cos\theta\, \delta^2 + \frac{\Delta \rho g \sin\theta}{4 \mu_v}\delta^2  \right),
\end{equation}

$\implies$

\begin{equation}
\frac{h'_{fg}\rho_v}{R}\frac{3 U \sin\theta}{4} \left( 1 + \frac{3 \rho_l U \cos\theta\delta^2}{4\mu_v R} + \frac{ \Delta \rho g \delta^2}{3 \mu_v U}   \right) ,
\end{equation}

$\implies$

\begin{equation}
\frac{h'_{fg}\rho_v}{R}\frac{3 U \sin\theta}{4} \left( 1 + \frac{3}{2}\frac{\rho_l}{\rho_v}\frac{\rho_v U D}{\mu_v}\left(\frac{\delta}{D}\right)^2\cos\theta + \frac{\Delta \rho g}{3U}\frac{\delta^2}{\mu_v}  \right),   
\end{equation}

$\implies$

\begin{equation}
\frac{h'_{fg}\rho_v}{R}\frac{3 U \sin\theta}{4} \left( 1 + \frac{3}{2}\frac{\rho_l}{\rho_v} Re_v\left(\frac{\delta}{D}\right)^2\cos\theta + \frac{\Delta \rho g}{3U}\frac{\delta^2}{\mu_v}  \right) ,  
\end{equation}

 
$\implies$
 
\begin{equation}
    \frac{h'_{fg}\rho_v}{R}\frac{3 U \sin\theta}{4} \left( 1 + \frac{3}{2}\frac{\rho_l}{\rho_v} Re_v\left(\frac{\delta}{D}\right)^2\cos\theta + \frac{Gr}{3Re_v}\left(\frac{\delta}{D} \right)^2  \right).
\end{equation}

Now, dividing the numerator and denominator by $\frac{h'_{fg}\rho_v}{R} \frac{3 U \sin\theta}{4} $, we can rewrite the denominator as

\begin{equation}
  1 + \frac{3}{2}\frac{\rho_l}{\rho_v} Re_v\left(\frac{\delta}{D}\right)^2\cos\theta + \frac{Gr}{3Re_v}\left(\frac{\delta}{D} \right)^2.
\end{equation}

Now, consider the numerator of equation \ref{diffdelta},

\begin{equation}
\begin{aligned}
\frac{1}{\frac{h'_{fg}\rho_v}{R} \frac{3 U \sin\theta}{4}} \frac{1}{D}\Bigg( \frac{k_v  \left(T_b - T_{sat}\right)}{\delta}\,+\, \sigma\epsilon(T^4_b - T^4_{sat})\,-\frac{k_l\Delta T_w \sin ^2 \theta }{\sqrt{\pi M\eta}}\,-\\ \frac{h'_{fg}\rho_v}{R}\left(\frac{3U \cos\theta \delta}{2} + \right.  \left. \frac{3 \rho_l U^2}{16\mu_v R}(3\cos^2\theta -1)\delta^3 + \frac{\Delta \rho g \cos\theta}{6\mu_v}\delta^3  \right) \Bigg).
\end{aligned}
\end{equation}


Term $\frac{1}{D}$ in the above expression comes from the non-dimensionalization of the left hand side of equation \ref{diffdelta} and $\frac{d\delta}{d\theta}$ is written as  $\frac{d(\delta/D)}{d\theta}D$.
We now non-dimensionalize all the terms of the above equation,
Consider the first term,

\begin{enumerate}
    \item 
    $$ \frac{1}{D}\frac{\frac{k_v \left(T_b-T_{sat}\right)  }{\delta}}{\frac{h'_{fg}\rho_v}{R} \frac{3 U \sin\theta}{4}}$$,
    
    $\implies$
    
    $$
      \frac{2 Cp_v (T_b - T_{sat})}{3h_{fg}} \frac{k_v}{Cp_v \rho_v U\sin\theta \delta} ,
    $$
    
    $\implies$
    
    $$\frac{2J_v}{3\frac{\rho_v Cp_v}{k_v}U D \frac{\delta}{D}\sin\theta}$$,
    
    $\implies$
    
    \begin{equation}
    \frac{2J_v}{3Pe_v\frac{\delta}{D}\sin\theta}.
    \end{equation}

Consider the second term,
    \item
    $$
      \frac{1}{D}\frac{q_r}{\frac{h'_{fg}\rho_v}{R} \frac{3 U \sin\theta}{4}}, 
    $$
    
    $\implies$
    
    \begin{equation}
    \frac{2q_r}{3\rho_v U h_{fg}\sin\theta}.  
    \end{equation}
  
Consider the third term,
   \item
   $$
      \frac{1}{D}\frac{\frac{k_l\Delta T_w \sin ^2 \theta }{\sqrt{\pi M\eta}}}{\frac{h'_{fg}\rho_v}{R} \frac{3 U \sin\theta}{4}},
   $$
   
   $\implies$
   
   $$
   \frac{2Cp_l (T_{sat}-T_w)}{3 h'_{fg}}\frac{k_l}{Cp_l\rho_v}\frac{\sin\theta}{\sqrt{\pi M \eta}}\frac{1}{U},
   $$
   
   $\implies$
   
   $$
    \frac{2}{3}J_l\alpha_l\frac{\rho_l}{\rho_v}\frac{\sin\theta}{\sqrt{\pi M \eta}}\frac{1}{U}.
   $$
   
   Substituting the values of $M$ and $\eta$ we get,

   $\implies$
   
   \begin{equation}
    \frac{2}{3}J_l\frac{\rho_l}{\rho_v}\sin\theta\frac{1}{\sqrt{\left( \frac{\pi Pe_l}{3} \left( \frac{2}{3}-\cos\theta+\frac{\cos ^3 \theta}{3} \right) \right)}}.  
   \end{equation}
   
Consider the fourth term,
\item
   $$
   \frac{1}{D}\frac{\frac{h'_{fg}\rho_v}{R}\frac{3U \cos\theta}{2}\delta}{\frac{h'_{fg}\rho_v}{R} \frac{3 U \sin\theta}{4}},
   $$
   
   $\implies$
   
   \begin{equation}
     2\cot\theta\left(\frac{\delta}{D}\right).  
   \end{equation}
 
 Consider the fifth term,
 \item
 $$
 \frac{1}{D}\frac{\frac{h'_{fg}\rho_v}{R}\frac{3 \rho_l U^2}{16\mu_v R}(3\cos^2\theta -1)\delta^3}{\frac{h'_{fg}\rho_v}{R} \frac{3 U \sin\theta}{4}},
 $$
 
 $\implies$
 
 $$ \frac{1}{2}\left(\frac{\delta}{D}\right)^3\frac{D}{\mu_v}\rho_l U \left(\frac{3\cos^2\theta-1}{\sin\theta} \right)$$,
 
 $\implies$

 $$
 \frac{1}{2}\left(\frac{\delta}{D}\right)^3\frac{\rho_l}{\rho_v} \frac{\rho_v U D}{\mu_v} \left(\frac{3\cos^2\theta-1}{\sin\theta} \right),
 $$
 
 $\implies$
 
 \begin{equation}
\frac{1}{2}\left( \frac{\delta}{D} \right)^3 \frac{\rho_l}{\rho_v}Re_v \left(\frac{3\cos^2\theta-1}{\sin\theta}\right).   
 \end{equation}
 
 Consider the sixth term,
 \item
 $$
 \frac{1}{D} \frac{\frac{\Delta \rho g \cos\theta}{6\mu_v}\delta^3}{\frac{h'_{fg}\rho_v}{R} \frac{3 U \sin\theta}{4}}, 
 $$  
 
 $\implies$
 
 $$
\frac{2}{9}\left( \frac{\delta}{D} \right)^3 D^2 \cot\theta \frac{\Delta \rho g}{U \mu_v},
 $$
 
 $\implies$
 
 \begin{equation}
 \frac{2}{9}\left( \frac{\delta}{D} \right)^3 \frac{Gr}{Re_v}\cot\theta.   
 \end{equation}
 
\end{enumerate}

 Arranging all the terms, we finally get,   

\begin{equation}
\begin{aligned}
\frac{d(\frac{\delta}{D})}{d\theta} = \frac{1}{1+ \frac{3\rho_l}{2\rho_v}Re_v (\frac{\delta}{D})^2 \cos\theta + \frac{1}{3}(\frac{\delta}{D})^2 \frac{G_r}{Re_v}} \left(\frac{2 J_v}{3 Pe_v\sin\theta(\frac{\delta}{D})}+\frac{2 q_r}{3 \rho_v U h'_{fg} \sin\theta} -\right. \\ \\ \left. 2\left(\frac{\delta}{D}\right)\cot\theta-\frac{1}{2} \frac{\rho_l}{\rho_v}Re_v \left(\frac{\delta}{D}\right)^3(\frac{3\cos^2\theta -1}{\sin\theta})-
\frac{2}{9}\frac{Gr}{Re_v}\left(\frac{\delta}{D}\right)^3 \cot\theta - \right. \\ \frac{2 \frac{\rho_l}{\rho_v} J_l \sin\theta}{3\left( \frac{\pi Pe_l}{3}\left(\frac{2}{3}-\cos\theta + \frac{\cos^3\theta}{3}  \right) \right)^\frac{1}{2} }  \Bigg).
\end{aligned}
\end{equation}
 
\textbf{Steps to solve expression \ref{eqinitial5}:}\\
The last term in equation \ref{eqinitial4} has to be solved separately as follow,

\begin{equation}
     \lim_{\theta\to 0} \left. \frac{2 \frac{\rho_l}{\rho_v} J_l \sin ^2\theta \frac{\delta}{D}}{3\left( \frac{\pi Pe_l}{3}\left(\frac{2}{3}-\cos\theta + \frac{\cos^3\theta}{3}  \right)    \right)^\frac{1}{2} }\right.
\label{lastterm1}
\end{equation}

The above equation can be written as 
\begin{equation}
     \lim_{\theta\to 0} \left. \frac{2 \frac{\rho_l}{\rho_v} J_l \sin ^2\theta \frac{\delta}{D}}{3\left( \frac{\pi Pe_l}{9}\left(2- 3\cos\theta + \cos^3\theta  \right)    \right)^\frac{1}{2} }\right.
\label{lastterm2}
\end{equation}

$\implies$

\begin{equation}
    2 \frac{\rho_l}{\rho_v} \frac{J_l}{\sqrt{\pi Pe_l}}\frac{\delta}{D}
     \lim_{\theta\to 0} \left. \frac{ \sin ^2\theta }{\left(2- 3\cos\theta + \cos^3\theta     \right)^\frac{1}{2} }\right.
\label{lastterm3}
\end{equation}

Now, consider the denominator inside the limit in equation \ref{lastterm3}. It can be written as,$$\left(1-\cos\theta \right)\left( 1-\cos\theta \right) \left( 2 + \cos \theta \right)$$
$\implies$

$$ 4 \sin ^4 \frac{\theta}{2} \left(2 + \cos\theta\right)$$

The numerator inside the limit in equation \ref{lastterm3} can be written as,
$$
\sin^2 \theta  = 4 \sin^2\frac{\theta}{2}\cos^2\frac{\theta}{2}
$$

Therefore, equation \ref{lastterm3} can be written as,
\begin{equation}
    2 \frac{\rho_l}{\rho_v} \frac{J_l}{\sqrt{\pi Pe_l}}\frac{\delta}{D}
     \lim_{\theta\to 0} \left. \frac{  4 \sin^2\frac{\theta}{2}\cos^2\frac{\theta}{2} }{\left(4 \sin ^4 \frac{\theta}{2} \left(2 + \cos\theta\right)\right)^\frac{1}{2} }\right.
\label{lastterm4}
\end{equation}
$\implies$

\begin{equation}
    2 \frac{\rho_l}{\rho_v} \frac{J_l}{\sqrt{\pi Pe_l}}\frac{\delta}{D}
     \lim_{\theta\to 0} \left. \frac{  4 \sin^2\frac{\theta}{2}\cos^2\frac{\theta}{2} }{2 \sin ^2 \frac{\theta}{2}\left(2 + \cos\theta\right)^\frac{1}{2} }\right.
\label{lastterm5}
\end{equation}

\begin{equation}
    2 \frac{\rho_l}{\rho_v} \frac{J_l}{\sqrt{\pi Pe_l}}\frac{\delta}{D}
     \lim_{\theta\to 0} \left. \frac{ 2 \cos^2\frac{\theta}{2}}{\left(2 + \cos\theta\right)^\frac{1}{2} }\right.
\label{lastterm6}
\end{equation}

Substituting the limit we get,

\begin{equation}
     \frac{4}{\sqrt {3 \pi Pe_l}} \frac{\rho_l}{\rho_v}J_l \frac{\delta}{D}
\label{lastterm7}
\end{equation}

Therefore, we can finally write,
\begin{equation}
     \lim_{\theta\to 0} \left. \frac{2 \frac{\rho_l}{\rho_v} J_l \sin ^2\theta \frac{\delta}{D}}{3\left( \frac{\pi Pe_l}{3}\left(\frac{2}{3}-\cos\theta + \frac{\cos^3\theta}{3}  \right)    \right)^\frac{1}{2} }\right|_\theta= - \frac{4}{\sqrt {3 \pi Pe_l}} \frac{\rho_l}{\rho_v}J_l \frac{\delta}{D}
\end{equation}

\end{document}